\documentclass[prb,twocolumn,superscriptaddress,floatfix,amsmath,amssymb,nofootinbib]{revtex4-2} 
\usepackage{graphicx}
\usepackage{hyperref}
\usepackage[capitalise]{cleveref}

\AtBeginDocument{
	\renewcommand{\Re}{\operatorname{Re}}
	\renewcommand{\Im}{\operatorname{Im}}
}

\DeclareMathOperator{\sgn}{sgn}

\begin{document}
\title{Applicability of Eliashberg theory for systems with electron-phonon and electron-electron interaction:
    a comparative analysis}
\author{Shang-Shun Zhang}
\author{Zachary M. Raines}
\author{Andrey V.Chubukov}
\affiliation{School of Physics and Astronomy and William I. Fine Theoretical Physics Institute, University of Minnesota, Minneapolis, MN 55455, USA}

\begin{abstract}
We present a comparative analysis of the validity of Eliashberg theory for the cases of fermions interacting with an Einstein phonon and with soft nematic fluctuations near an Ising-nematic/Ising-ferromagnetic quantum-critical point (QCP).
In both cases, Eliashberg theory is obtained by neglecting vertex corrections.
For the phonon case, the reasoning to neglect
 vertex corrections is the Migdal ``fast electron/slow boson'' argument because the phonon velocity is much smaller than the Fermi velocity, $v_F$.
The same argument allows one to compute the fermionic self-energy within Eliashberg theory perturbatively rather than self-consistently.
For the nematic case, the velocity of a collective boson is comparable to $v_F$ and this argument apparently does not work.
Nonetheless,
we argue that
while
 two-loop vertex corrections near a nematic QCP are not small parametrically,
they are small numerically.
At the same time, perturbative calculation of the fermionic self-energy can be rigorously justified when the fermion-boson coupling is small compared to the Fermi energy.
Furthermore, we argue that for the electron-phonon case Eliashberg theory breaks down at some distance from where the dressed Debye frequency would vanish, while for the nematic case it holds all the way to a QCP.\@
From this perspective, Eliashberg theory for the nematic case actually works better than for the electron-phonon case.
\end{abstract}

\maketitle

\section{Introduction}

Migdal-Eliashberg (ME) theory has been developed to describe electron-phonon interaction in the normal state and phonon-induced superconductivity~\cite{migdal1958interaction,eliashberg1960interactions}.
Over the years, it has been successfully applied to numerous electron-phonon systems~\cite{Allen_1975,Karakozov_1976,Marsiglio_91,combescot, marsiglio2020eliashberg,chubukov2020eliashberg}.
The theory, as formulated by Eliashberg~\cite{eliashberg1960interactions},
consists of a set of three coupled self-consistent one-loop equations for the normal and anomalous fermionic self-energies and the phonon polarization.
 These equations can be derived either diagrammatically or from the variational Luttinger-Ward functional~\cite{lw}, assuming that vertex corrections can be neglected.
ME argued that vertex corrections are small in the ratio of the Debye frequency $\omega_D$ to the Fermi energy $E_F$.
The physical reasoning, due to Migdal~\cite{migdal1958interaction}, is that in processes leading to vertex corrections, a fermion is forced to vibrate at the phonon frequency, which for small $\omega_D/E_F$ is far from its own resonance.
In mathematical terms, the strength of vertex corrections is governed by the dimensionless Eliashberg parameter $\lambda_E = g^2/(\omega_D E_F)$. This parameter is different from the dimensionless coupling $\lambda = g^2/\omega^2_D$, which determines mass renormalization: $\lambda_E = \lambda (\omega_D/E_F) \ll \lambda$.
By this reasoning, the one-loop theory remains under control not only at weak coupling, where $\lambda <1$, but also at strong coupling, where $\lambda >1$ but $\lambda_E <1$ (i.e., at $\omega_D < g < (\omega_D E_F)^{1/2}$).
Still, when the Debye frequency softens, $\lambda_E$ eventually becomes large and the one-loop description breaks down.
At larger $\lambda_E$, higher loop processes become relevant and eventually lead to polaronic description.

In recent years, a similar description has been applied to metals in which electron-electron interactions (screened Coulomb repulsion) give rise to long-range particle-hole order
either in the spin or charge channel~\cite{Millis1992,Altshuler1995a,Sachdev1995,acs,*finger_2001,acf,*Abanov2000,*acs2,*acn,Subir,Subir2,nick_b,paper_1,wang,max2,vojta,
    efetov,*efetov2,*efetov3,tsvelik,Bauer2015,ital,*ital2,*ital3,wang23,wang_22,triplet,*triplet2,*triplet3,sslee2, steve_sam,PALee1989,*Monien1993,*Nayak1994,*Altshuler1994,*Kim_1994,rech_2006,DellAnna2006,*Metzner2003,Maslov2010,Fradkin2016,Esterlis2021,
    10.21468/SciPostPhys.13.5.102,Klein2018,*Klein2019,*Klein2020,*Klein2022,Nosov2023,Zhang2023,Foster2023,*Nosov2024}.
Near the onset of such an order, i.e., near a quantum-critical point (QCP), it is natural to assume that the low-energy physics is described by an effective model with fermion-fermion interactions mediated by soft fluctuations of the bosonic order parameter that condenses at the QCP.\@
The theory near a QCP is based on the same set of coupled equations that Eliashberg obtained for the electron-phonon case and bears his name.

The validity of Eliashberg theory for a system near a QCP has been questioned, however, on the grounds that
the ``fast fermion/slow boson'' argument, used to justify the neglect of vertex corrections, is not applicable anymore because collective excitations are made out of fermions, and their velocity is of order $v_F$ (see e.g., Ref.~\onlinecite{shi2024excitonic}).
The counter-argument~\cite{paper_1} is that soft collective excitations are Landau overdamped and for this reason do behave as slow modes compared to fermions, i.e., Migdal's reasoning is still valid, albeit for a different reason.

The aim of this communication is to settle the issue of applicability of the Eliashberg (one-loop) theory for a quantum-critical metal near a QCP.\@
We compare the criteria for applicability of the Eliashberg theory for fermions interacting with a soft Einstein phonon with $\omega_D \ll E_F$, and with soft nematic fluctuations near an Ising-nematic/Ising-ferromagnetic  QCP.\@
We argue that the criteria for these two cases are somewhat similar, but not identical.

Below we restrict our analysis to the normal state at $T=0$.
This is a putative ground state as the true one is a superconductor in both cases.
However, for spin-singlet pairing, the criterion for validity of the Eliashberg theory for a superconductor is actually weaker than that for the normal state as singular quantum corrections to the self-energy cancel in the gap equation like contributions from non-magnetic impurities~\cite{acn,paper_4,paper_5}.
For this reason, if ME theory is applicable in the putative normal state at $T=0$, it is also applicable to a superconductor.
Restriction to $T=0$ is essential because thermal fluctuations at a finite $T$ do not fit into ME reasoning and may destroy ME theory faster than quantum fluctuations
(see Refs.~\onlinecite{chubukov2020eliashberg,yuz_2} for more detail).

We also do not discuss here logarithmic singularities in the fermionic self-energy in the Ising-nematic/Ising-ferromagnetic case that emerge at three-loop and higher-orders~\cite{metlitski2010quantum,Lee2009,Lee2018,Tobias2015,pimenov}.
Most likely, these corrections generate an anomalous dimension for the fermionic propagator but the branch cut in the Green's function remains at the same place as in the Eliashberg theory.

Finally, for the electron-phonon case, we do not discuss here the proposals~\cite{Alexandrov1994,Millis1996,x_representation,sous2023bipolaronic}
that the transition to a ground state with polarons may be first order and happen already at $\lambda = O(1)$
due to singular behavior of
$n$-loop
contributions to the self-energy with $n \gg 1$ (Ref. \cite{sous2023bipolaronic}).
In this communication we restrict with the analysis of the self-energy at the two-loop order.

\subsection{Summary of the results}

The summary of our results is the following.
As said above, for the electron-phonon case, the smallness of two-loop vertex corrections is controlled by the Eliashberg parameter
$\lambda_E = g^2/(\omega_D E_F) = \lambda \omega_D/E_F$.
This parameter is small at weak coupling, when $\lambda <1$, but remains small also at $\lambda >1$, as long as $\lambda < E_F/\omega_D$.
At larger $\lambda$ vertex corrections become parametrically large and Eliashberg theory breaks down.
This happens at a finite $\omega_D$, unless one takes the double limit $E_{F}\to\infty$ and $\omega_{D}\to 0$ while maintaining $g^2 \ll \omega_D E_{F}$ ~\cite{paper_5,zhang_22}.
If $E_F$ is kept finite, as we do in this paper, $\lambda_E = O(1)$ is reached at a finite $\omega_D$.

For the Ising-nematic/Ising-ferromagnetic case, vertex corrections are small at weak coupling, when the corresponding $\lambda^*$,
defined via
$\lambda^{*} = \lim_{\omega_m \to 0} \partial \Sigma/ \partial \omega_m$, 
is small.
In the strong coupling regime $\lambda^* >1$, vertex corrections are not small parametrically, however,
they remain of order one and are small numerically (of order $10^{-2}$).
This behavior holds all the way up to the QCP, where vertex corrections remain numerically
 small.

There is more.
The two Eliashberg equations for the dynamical fermionic self-energy and bosonic polarization, obtained from the Luttinger-Ward functional, are coupled self-consistent one-loop equations.
The one for the bosonic polarization accounts for Landau damping of the
 bosons.
For the electron-phonon case the \emph{same} parameter $\lambda_E <1$ that justifies the neglect of vertex corrections also allows one to simplify these equations in the following way: neglect the Landau damping and replace the self-consistent one-loop equation for the fermionic self-energy by the perturbative one-loop formula, in which the self-energy is expressed in terms of the bare fermionic propagator.
This simplification holds because, even at strong coupling ($\lambda >1$), typical fermionic energies $\epsilon_k$ are of order $\lambda_E E_F \ll E_F$, in which case integration over $\epsilon_k$ can be extended to infinite limits, and $\int d \epsilon_k/(i(\omega_m + \Sigma (\omega_m))-\epsilon_k) = -i\pi \sgn \omega_m$, like for free fermions.
The ability to replace the self-consistent equation for the self-energy by a perturbative one can be also expressed as a consequence of the smallness of typical fermionic momenta transverse to the Fermi surface compared to typical momenta along the Fermi surface
($\lambda_E k_F$
vs. \ $k_F$).

For the Ising-nematic/Ising-ferromagnetic case Landau damping is relevant but the self-consistent one-loop Eliashberg equations can be still reduced to the perturbative ones because, like in the phonon case, typical momenta transverse to the Fermi surface are parametrically smaller than typical momenta along the Fermi surface.
The smallness holds in $\lambda^*_E$, which is the ratio of the fermion-boson coupling and the Fermi energy.
This ratio must be small, as otherwise the low-energy description would not be valid.
The value of $\lambda^*_E$ does not depend on the distance to the QCP, and hence for the Ising-nematic/Ising-ferromagnetic case the one-loop perturbation theory remains valid even at the QCP.\@

The outcome of this analysis is that Eliashberg theory for the electron-phonon case is rigorously justified for $\lambda_E <1$, even when the coupling $\lambda >1$.
By the same condition $\lambda_E <1$, the fermionic self-energy can be computed in one-loop perturbation theory rather than self-consistently.
Because $\lambda_E$ is
inversely
proportional to  $\omega_D$, the theory is valid only at $\omega_D$ above some critical value.

For the Ising-nematic/Ising-ferromagnetic case, Eliashberg theory is not rigorously justified at strong coupling, but vertex corrections remain $O(1)$ and are small numerically.
At the same time, there exists another small parameter $\lambda^*_E$ (the ratio of the interaction to the Fermi energy), which justifies the computation of the fermionic self-energy in a perturbative one-loop analysis rather than self-consistently.
This parameter remains small even at a QCP, i.e., Eliashberg theory can be applied to a system at a QCP, if one is willing to neglect numerically small vertex corrections.

From this last perspective, Eliashberg theory near a nematic QCP
works better than for fermions interacting with a soft Einstein phonon, despite that the velocity of collective nematic fluctuations is of order $v_F$.

The structure of the paper is the following. In the next section we briefly review Eliashberg theory for
electron-phonon interaction~\cite{eliashberg1960interactions,marsiglio2020eliashberg,Chubukov_2020b}
and discuss the strength of one-loop vertex corrections.
In \cref{sec:qcm} we discuss fermions near a nematic QCP.\@
In \cref{sec:comp} we compare the two cases and present our conclusions.

\section{Eliashberg theory for electron-phonon system}
\label{sec:el-ph}

For definiteness we consider interaction with an Einstein phonon.
Eliashberg theory for the normal state
consists of the set of two equations for the fermionic self-energy $\Sigma (k)$ and
phonon polarization $\Pi (q)$, where
$k \equiv (\mathbf{k}, \omega_m)$ and $q \equiv (\mathbf{q}, \Omega_n)$.
The equations look simplest along the Matsubara axis, when $\omega =\omega_m = (2m+1)\pi T$ and
$\Omega = \Omega_n = 2n\pi T$.
 The full fermionic Green's function and the full phonon propagator are related to $\Sigma$ and $\Pi$ via
\begin{equation}
    G^{-1} (k) = G^{-1}_0 (k) +
    i
    \Sigma (k),\ \chi^{-1} (q) = \chi^{-1}_0 (q) + \Pi (q),
    \label{eq:a1}
\end{equation}
where $G_0$ and $\chi_0$ are electron and phonon propagators in the absence of interaction
\begin{equation}
    G_0(k) = \frac{1}{i\omega_{m} - \epsilon_\mathbf{k}},\quad \chi_0 (q) = \frac{\chi_0}{\Omega_{n}^2+ \omega^2_D}.
    \label{eq:bare-gfs}
\end{equation}
To simplify calculations, we assume a parabolic dispersion, $\epsilon_\mathbf{k} = (\mathbf{k}^2 -k^2_F)/(2m)$.
Note that we define the Matsubara self-energy as $i \Sigma (\omega_m)$
(the ${\bf k}$-dependent term in $\Sigma(k)$ is proportional to $\lambda_E$ and thus negligible at small $\lambda_E$, see below),
so that the inverse electron propagator can be written $G^{-1}_{0} = i(\omega_{m} + \Sigma(\omega_{m})) - \epsilon_{\mathbf{k}}$.
At $T=0$, which we consider here, both $\omega_m$ and $\Omega_n$ are continuous variables.

\begin{figure}
    \centering
    \includegraphics[width=0.9\linewidth]{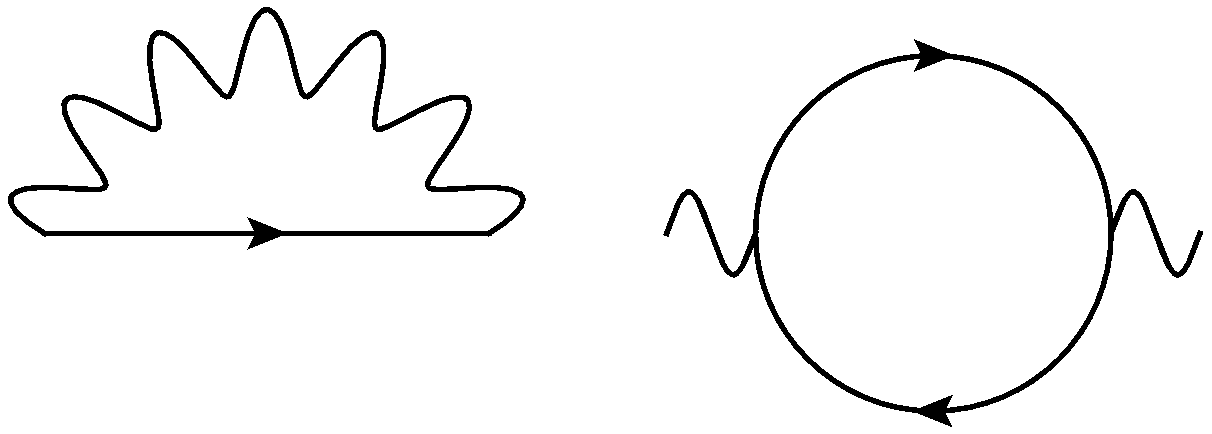}
    \caption{Self-consistent one-loop electron self-energy (left panel) and polarization bubble for Einstein phonon (right panel).
        Solid lines indicate the renormalized electronic Green's function $G(k)$ and wavy lines the renormalized Einstein phonon propagator $\chi(q)$.\label{fig:selfE_1loop}}
\end{figure}

The self-consistent equations for $\Sigma$ and $\Pi$ can be either derived diagrammatically, see \cref{fig:selfE_1loop}, or obtained as stationary solutions of the Luttinger-Ward functional.
In analytic form, they are
\begin{gather}
    \Sigma(k) =
    i\tilde{g}^2 \int \frac{d^2\mathbf{q} \, d\Omega_n}{(2\pi)^3} G(k+q) \chi(q), \label{eq:sigma}\\
    \Pi (q) = 2 \tilde{g}^2 \int \frac{d^2\mathbf{k} \, d\omega_m}{(2\pi)^3} G(k+q)G(k) \label{eq:pi}
\end{gather}
where $\tilde{g}$ is the
electron-phonon
coupling.

There are three energy scales in these equations: the Debye frequency $\omega_D$, the Fermi energy $E_F = k^2_F/(2m)$, and the dimension-full effective interaction $g = (\tilde{g}^2 N_{F}\chi_0)^{1/2}$, where
$N_{F}=k_F/(2\pi v_F)$
is the density of states at the Fermi level
per spin component.
 Out of these one can introduce two dimensionless ratios
\begin{equation}
    \lambda = \frac{g^2}{\omega^2_D},\quad \lambda_E = \frac{g^2}{E_F \omega_D} = \lambda \frac{\omega_D}{E_F}.
    \label{eq:a3}
\end{equation}
Eliashberg theory
is constructed under the assumptions that $E_F$ is the largest energy scale and $\lambda_E \ll \lambda$.
The ME argument
is that the strength of vertex corrections is determined by the smaller
 $\lambda_E$, while fermionic mass renormalization is determined by the larger
$\lambda$.
When both $\lambda $ and $\lambda_E$ are small, the theory falls into the weak coupling limit with $G$ and $\chi$ close to their bare expressions.
In the regime $\lambda >1$, $\lambda_E <1$, mass renormalization is large and the self-energy is larger than bare $\omega_{m}$ over a wide range of frequencies, yet
 vertex corrections are still small.

We will be chiefly interested in the strong coupling regime $\lambda >1$, $\lambda_E <1$.
We go beyond previous work~\cite{chubukov2020eliashberg} and
analyze vertex corrections in 2D for all phonon momenta. We also
compute the two loop self-energy with vertex correction included. We show that the latter is small in $\lambda_E$, and
becomes $O(1)$ when $\lambda_E \sim 1$.

ME found~\cite{migdal1958interaction,eliashberg1960interactions}
that the same small parameter $\lambda_E$ allows one to simplify
calculations within Eliashberg theory and obtain a simple expression for the self-energy,
$\Sigma^{(E)}(\omega_m) = \lambda \omega_D \arctan \left(\omega_m/\omega_D\right)$ (\cref{eq:sigma_eli} below), which depends only on frequency.
Here we analyze the corrections to this expression, both analytically and numerically.
We show that \cref{eq:sigma_eli} can be rigorously justified only for frequencies $\omega_m \ll E_F$.
At $\omega_m \sim E_F$, the expression is more complex
(\cref{eq:delta-sigma-large-frequency} below).
 In particular, $\Sigma (k_F, \omega_m)$ becomes complex on the Matsubara axis~\footnote{This does not violate
 Kramers-Kronig relations, but when $\Sigma (k_F, \omega_m)$ is complex, both $\Sigma^\prime (k_F, \omega)$ and $\Sigma^{\prime\prime} (k_F, \omega)$ on the real axis have even and odd components in $\omega$.},
similar to the self-energy in SYK-type models~\cite{PhysRevB.63.134406,Wang2020}.

We also show that the momentum-dependent part of the self-energy remains small as long at $\lambda_E <1$.

We discuss the solution of the Eliashberg equations first and then use it to analyze the strength of vertex corrections.

\subsection{Solution of Eliashberg equations}

The equations for $\Sigma (k)$ and $\Pi (q)$ are coupled and in principle have to be solved together.
We argue, however, that for $\lambda_E \ll 1$, the two equations can be solved independently.
To demonstrate this, we make two assumptions and verify both a posteriori.
First, we assume that
$\partial \Sigma/\partial \omega_m$ 
scales as $\lambda$ and is large at strong coupling, while
$\partial \Sigma (k)/\partial \epsilon_k$
scales as $\lambda_E$ and is small when $\lambda_E <1$.
Accordingly, we approximate $\Sigma (\mathbf{k}, \omega_m)$ by $\Sigma (\omega_m)$.
Second, we assume that $\Sigma (\omega_m)$ is parametrically smaller than $E_F$.

Consider
 \cref{eq:pi}
for $\Pi (q)$ first.
We express $\Pi (q) = \Pi (\mathbf{q}, \Omega_n)$
as a sum of static and dynamic pieces
$\Pi (\mathbf{q}, 0) + \delta \Pi (\mathbf{q}, \Omega_n)$.
The static part $\Pi (\mathbf{q}, 0)$ generally comes from fermions with energies of order $E_F$.
It renormalizes $\omega_D$ and $\chi_0$, and may also give rise to a momentum dependence of the phonon propagator.
For the purposes of this study we assume that this renormalization is already included into $\chi_0 (q)$ and that
$\chi_0 (q)$ is given by \cref{eq:bare-gfs} (for more discussions on this issue see Refs.~
[\onlinecite{marsiglio2021phonon,chubukov2020eliashberg}]).
The second, dynamical, term comes from fermions with low energies.
For this term we obtain, after integrating over the angle between phonon $\mathbf{q}$ and fermionic $\mathbf{k}$:
\begin{widetext}
\begin{equation}
    \delta \Pi (q) = -2i \tilde{g}^2\int \frac{d \omega_m}{2\pi} \int \frac{kdk}{2\pi} \frac{1}{i \tilde{\Sigma} (\omega_m) - \epsilon_k} \left[\frac{\sgn (\omega_m + \Omega_n)}{\sqrt{\frac{(kq)^2}{m^2} + \left({\tilde \Sigma} (\omega_m + \Omega_n) +i\left(\epsilon_k + \frac{q^2}{2m}\right)\right)^2}} - {\text{the same at}}~ \Omega_n =0\right],
    \label{eq:delta-pi-angle-integrated}
\end{equation}
where for brevity we have defined
$\tilde{\Sigma} (\omega_m) = \omega_m + \Sigma (\omega_m)$.
For $\omega_m, \Omega_n \ll E_F$ and a generic $q \sim k_F$, the terms $\tilde{\Sigma} (\omega_m + \Omega_n)$ and $\epsilon_k$ under the square-root in \cref{eq:delta-pi-angle-integrated} are much smaller than $E_F$ and can be neglected compared to the two $q-$dependent terms, which are of order $E_F$.
\Cref{eq:delta-pi-angle-integrated} then simplifies to
\begin{equation}
    \delta \Pi (q) = -2i \tilde{g}^2 N_F \int \frac{d \omega_m}{2\pi} \int \frac{d\epsilon_k}{i \tilde{\Sigma} (\omega_m) - \epsilon_k} \frac{\sgn (\omega_m + \Omega_n) - \sgn (\omega_m)} {v_F q \sqrt{1- \left(\frac{q}{2k_F}\right)^2}}
\end{equation}
\end{widetext}
where $N_{F}$ is the density of states at the Fermi level.
Integration over $\epsilon_k$ can now be extended to infinite limits
(up to terms of order $\tilde{\Sigma}(\omega_{m})/E_{F}\ll 1$)
and yields
$\int d \epsilon_k/(i \tilde{\Sigma} (\omega_m) - \epsilon_k) = -i \pi \sgn (\omega_m)$.
Integrating then over frequency, we obtain
\begin{equation}
    \delta \Pi (q) = 2\tilde{g}^2 N_F\frac{\Omega_n}{v_F q \sqrt{1- \left(\frac{q}{2k_F}\right)^2}}.
    \label{eq:delta-pi}
\end{equation}
This expression is valid for $q <2k_F$, which is the largest momentum transfer on the Fermi surface.
At small $q$, it reduces to conventional Landau damping. Substituting $\delta \Pi$ into \cref{eq:a1} and using \cref{eq:bare-gfs}, we obtain
\begin{equation}
    \chi (\mathbf{q}, \Omega_n) = \frac{\chi_0}{\Omega^2_m + \omega^2_D + \gamma \frac{|\Omega_n|}{q \sqrt{1- \left(\frac{q}{2k_F}\right)^2}}},
    \label{eq:chi-phonon-renormalized}
\end{equation}
where $\gamma = 2 g^2/v_F$ and
$g^2 = \tilde{g}^2 N_F \chi_0$.

\begin{figure}
    \centering
    \includegraphics[width=0.9\linewidth]{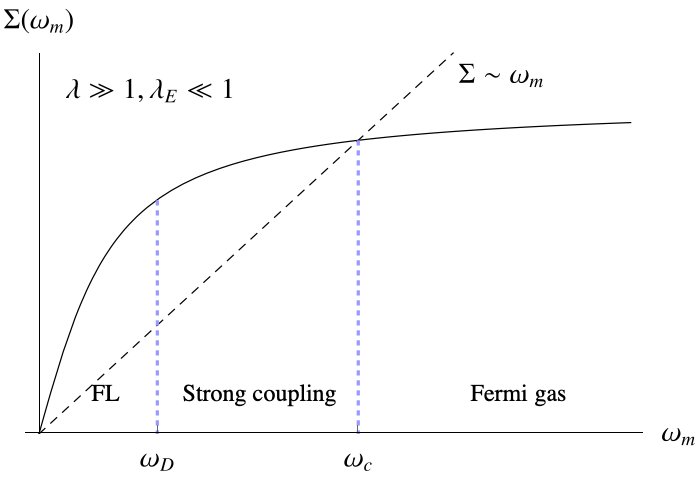}
    \caption{(Color online)
        One-loop perturbative
        Eliashberg self-energy $\Sigma^{(E)}(\omega_m)$ at $\lambda \gg 1$, but $\lambda_E \ll 1$.
        The self-energy is linear in $\omega_m$ for $\omega_m < \omega_D$ and saturates at larger $\omega_m$.
        It is larger than the bare $\omega_m$ for $\omega_m < \omega_c = (2/\pi) E_F \lambda_E$.
        For $\omega_m < \omega_D$, the system is in the Fermi liquid regime.
        In between $\omega_D$ and $\omega_C$, it displays strong coupling, non-Fermi liquid behavior.
        At larger $\omega_m$ the self-energy is smaller than $\omega_m$ and the system behaves as a Fermi gas.
        \label{fig:schematic_elph}
    }
\end{figure}

It is convenient to measure $q$ in units of $2k_F$ and $\Omega_n$ in units of $\omega_D$.
Introducing $\bar{q} = q/(2k_F)$ and ${\bar \Omega}_m = \Omega_n/\omega_D$, we re-express \cref{eq:chi-phonon-renormalized} as
\begin{equation}
    \chi (\mathbf{q}, \Omega_n) = \frac{\chi_0}{\omega^2_D} \frac{1}{1 + {\bar \Omega}^2_m + \frac{\lambda_E}{2}
    \frac{|{\bar \Omega}_m|}{\bar{q}\sqrt{1- \bar{q}^2}}}
    \label{eq:a9}
\end{equation}
We see that the Landau damping term contains $\lambda_E$ in the prefactor and is therefore small compared to one of the two other
terms for all values of ${\bar \Omega}_m$ provided $\bar{q}$ is not close to either zero or one.

We now substitute $\chi (\mathbf{q}, \Omega_n)$ from \cref{eq:a9} into \cref{eq:sigma} for the self-energy.
We compute separately $\Sigma (\mathbf{k}_{F}, \omega_m)$ and $\Sigma (\mathbf{k}, 0)$. The calculation of $\Sigma (\mathbf{k}_{F}, \omega_m)$ parallels the one for $\delta \Pi$: we first integrate over the angle between $\mathbf{k}_F$ and $\mathbf{q}$ and obtain
\begin{widetext}
\begin{equation}
    \Sigma (\omega_m) =\frac{g^2}{N_F} \int \frac{d \Omega_n}{2\pi} \int_0^{2k_F} \frac{qdq}{2\pi}
    \frac{1}{\Omega^2_m + \omega^2_D + \gamma \frac{|\Omega_n|}{q\sqrt{1-(q/2k_F)^2}}}~
    \frac{\sgn (\omega_m + \Omega_n)}{\sqrt{(v_F q)^2 + \left(\tilde{\Sigma} (\omega_m + \Omega_n) +i \frac{q^2}{2m}\right)^{2}}}.
    \label{eq:sigma-phonon-angle-integrated}
\end{equation}
\end{widetext}
We set $\omega_m \ll E_F$ and assume and verify a posteriori that
(i) typical $q$ are of order $k_F$ and typical $\Omega_n$ are of order $\omega_m$ and (ii) $\tilde{\Sigma} (\omega_m + \Omega_n)$ is parametrically smaller than $E_F$. We then evaluate the integrals over $q$ and $\Omega_q$ neglecting the Landau damping term and ${\tilde \Sigma} (\omega_m + \Omega_n)$ under the square root.
 The calculation is elementary, and we obtain
\begin{equation}
    \Sigma^{(E)}(\omega_m) = \lambda \omega_D \arctan \left( \frac{\omega_m}{\omega_D} \right),
    \label{eq:sigma_eli}
\end{equation}
where the index $E$ denotes that this is the leading self-energy in Eliashberg theory.
The self-energy behaves as $\lambda \omega_m$ at small frequencies and saturates at $(\pi/2) \lambda \omega_D$ at higher frequencies.
We emphasize that the right-hand side of \cref{eq:sigma_eli} does not depend on the self-energy and has the same form as if we used the Green's function for free fermions.
The same holds for $\delta \Pi (q)$. In other words, to this accuracy, the one-loop self-consistent theory
becomes equivalent to the one-loop perturbation theory. We plot $\Sigma^{(E)}(\omega_m)$ in \cref{fig:schematic_elph}.

We next
check
 the accuracy of approximations used to obtain \cref{eq:sigma_eli}.
The typical $q$ for the momentum integral are of order $k_F$ and typical $\Omega_n$ for the frequency integral are of order $\omega_m$, as we assumed.
Further, the prefactor $\lambda \omega_D$ in \cref{eq:sigma_eli} can be equivalently expressed as $\lambda_E E_F$. We see that for small $\lambda_E$,
$\tilde{\Sigma} (\omega_m + \Omega_n) \ll E_F$. This justifies neglecting the self-energy term in \cref{eq:sigma-phonon-angle-integrated} and in the calculation of $\delta \Pi (q)$. Also, the self-energy $\Sigma^{(E)}(\omega_m)$ exceeds the bare $\omega_m$ at small $\omega_m$, up to $\omega_c =
    (\pi/2) \lambda \omega_D = (\pi/2)
    \lambda_E E_F$. As long as $\lambda_E$ is small, $\omega_c$ is parametrically smaller than $E_F$.

\subsection{Subleading terms in the self-energy}
\label{sec:subleading_elph}

We now go beyond the leading term and estimate the subleading terms in the self-energy.
They are the corrections to the dynamic piece of the self-energy,
arising from
 (i) keeping either the Landau damping term
or (ii) keeping $\tilde{\Sigma}$ in right-hand-side of \cref{eq:sigma-phonon-angle-integrated},
and (iii) the correction from the static piece of the self-energy $\Sigma(\mathbf{k}, 0)$.

Keeping the Landau damping term in \cref{eq:sigma-phonon-angle-integrated} leads to
a relative correction $(1+X (\omega_m))$, which is the largest at $\omega_m \sim \omega_D$, where $X \sim \lambda_E \log^2 \lambda_E$.
This is a small correction when $\lambda_E \ll 1$. The leading correction from keeping
$\tilde{\Sigma} (\omega_m + \Omega_n)$ in the integrand in \cref{eq:sigma-phonon-angle-integrated} comes from $q \approx 2k_F$.
This correction, which we label as $\delta \Sigma (\omega_m)$, is a bit tricky. First, there is a non-zero
$\delta \Sigma (0)$. This contribution is purely imaginary in our notation of Matsubara self-energy as $i\Sigma (\omega_m)$, i.e., it is the real part of the actual self-energy.
This term is often absorbed into the renormalization of the bare chemical potential. This is acceptable for the low-energy description, but we emphasize that
$\Im[\delta \Sigma (\omega_m)]$ generally depends on frequency.
For this reason we don't subtract $\delta \Sigma (0)$ from $\delta \Sigma (\omega_m)$. Second, at $|\omega_m| < \omega_D$, the real part of $\delta \Sigma (\omega_m)$ is linear in $\omega_m$, like $\Sigma^{(E)} (\omega_m)$, but with the prefactor $\lambda \sqrt{\lambda_E}$, which is
smaller than that in \cref{eq:sigma_eli} by $\sqrt{\lambda_{E}}$.
For $\omega_D < |\omega_m| < E_F$, this correction becomes
\begin{equation}
    \delta \Sigma (\omega_m) = -\omega_c
    \left(\frac{|\tilde{\Sigma}^{(E)} (\omega_m)|}{E_F}\right)^{1/2} \frac{1 + i \sgn \omega_m}{\pi}.
    \label{eq:delta-sigma-large-frequency}
\end{equation}
Substituting $|\tilde{\Sigma}^{(E)}
    (\omega_m)| = |\omega_m| + \omega_c$, we find that for $|\omega_m| < \omega_c$ $\delta \Sigma (\omega_m)$ is smaller than $\Sigma^{(E)}$ by
    $(\omega_c/E_F)^{1/2} \sim (\lambda_E)^{1/2}$.
At larger frequencies, $\delta \Sigma$ increases and at $\omega_m \sim E_F$
becomes of the same order as $\Sigma^{(E)} (\omega_m)$.
We emphasize that this holds even when $\lambda_E$ is small.
At even larger $|\omega_m| > E_F$, the full self-energy has to be re-evaluated.
We will not analyze this frequency range here.

We note that $\delta \Sigma (\omega_m)$ is a complex function of Matsubara frequency.
Similar behavior of $\Sigma (\omega_m)$ has been previously reported in SYK-type systems~\cite{PhysRevB.63.134406,Wang2020}.
In our case the appearance of a complex $\Sigma (\omega_m)$ reflects the absence of particle-hole symmetry for a parabolic dispersion.

The computation of the static piece $\Sigma (\mathbf{k}, 0)$ proceeds in the same way.
After angular integration, the left-hand side of
\cref{eq:sigma} becomes
\begin{widetext}
\begin{equation}
    \Sigma (\mathbf{k}, 0) =\frac{g^2}{N_F} \int \frac{d \Omega_n}{2\pi} \int_0^{2k_F} \frac{qdq}{2\pi}
    \frac{1}{\Omega^2_m + \omega^2_D + \gamma \frac{|\Omega_n|}{q\sqrt{1-(q/2k_F)^2}}}
    \frac{\sgn (\Omega_n)}{\sqrt{(v_F q)^2 + \left(\tilde{\Sigma} (\Omega_n) +i \left(\epsilon_k + \frac{q^2}{2m}\right)\right)
            ^{2}}}.
    \label{eq:a12}
\end{equation}
Expanding in $\epsilon_k$, we obtain
\begin{equation}
    i\Sigma (\mathbf{k}, 0) = \epsilon_k \frac{2g^2 }{N_F} \int \frac{d \Omega_n}{2\pi} \int_0^{2k_F} \frac{qdq}{2\pi}
    \frac{1}{\Omega^2_m + \omega^2_D + \gamma \frac{|\Omega_n|}{q\sqrt{1-(q/2k_F)^2}}}
    \frac{|\tilde{\Sigma}(\Omega_n)|}{\left({\tilde \Sigma}^2 (\Omega_n) + (v_F q)^2 (1 - (q/(2k_F)^2))\right)^{3/2}}.
    \label{eq:a14}
\end{equation}
\end{widetext}
The largest contributions to the momentum integral comes from small $q$ and from $q \approx 2k_F$. Evaluating the contributions from these two regions and performing the subsequent integration over frequency, we obtain
\begin{equation}
    i\Sigma (\mathbf{k}, 0) = \lambda_E \epsilon_k.
    \label{eq:a15}
\end{equation}
As a consequence, the $\epsilon_k$ term in the bare Green's function gets multiplied by the factor $1-\lambda_E$. This renormalization is small when $\lambda_E$ is small.

Summarizing, we find that the condition $\lambda_E \ll 1$ allows us to (i) neglect the Landau damping term in the bosonic propagator, (ii) neglect the momentum dependence of the self-energy, and (iii) approximate $\Sigma (\omega_m)$ by
the one-loop perturbative result
\cref{eq:sigma_eli}. For the latter, the corrections, which make $\Sigma (\omega_m)$ complex, are parametrically small as long as $\omega_m$ remains below $\omega_c \sim \lambda_E E_F$. At larger frequencies, the corrections get stronger, and at $\omega_m \sim E_F$ become
of the same order as $\Sigma^{(E)} (\omega_m)$.
To the best of our knowledge, this last result has not been presented earlier.

The smallness of corrections to the one-loop perturbative
$\Sigma^{(E)} (\omega_m)$
can be understood by comparing relative energy scales and invoking the argument about slow phonons and fast electrons, which typically is reserved for vertex corrections. Indeed, the argument implies that for the same frequency, fermionic momenta are far smaller than phonon momenta. In our case, typical phonon momenta are of order $k_F$, while typical fermionic momenta are of order $\tilde{\Sigma} (\omega_m)/v_F$. For $|\omega_m| < \omega_c$, where \cref{eq:sigma_eli} for
$\Sigma^{(E)} (\omega_m)$
is rigorously justified,
typical fermionic momenta are smaller by $\lambda_E$.
In practical terms, this separation
allows one to approximate $\int d^2k$ in the Eliashberg formula for the self-energy for a generic $\chi (q)$ by $2\pi N_F \int d \epsilon_{k'} d \theta$, where $\mathbf{k}' =\mathbf{k} + \mathbf{q}$ and $\theta$ is the angle between $\mathbf{k}$ and $\mathbf{k}'$, with both set on the Fermi surface.
\cref{eq:sigma} then reduces to
\begin{widetext}
\begin{equation}\label{eq:sigma_decomposed}
    \Sigma(\mathbf{k}, \omega_m) = \Sigma (\omega_m) = i \tilde{g}^2 N_F \int{d\Omega_n \over 2\pi} \chi_L(\Omega_n)
    \int {d \epsilon_{\mathbf{k}\prime}} G(k^\prime, \omega_m + \Omega_n)
    = g^2 \int_0^{\omega_m} d\Omega_n \chi_L(\Omega_n),
\end{equation}
\end{widetext}
where we have defined $\chi_L(\Omega_{m}) = (1/2\pi) \int d \theta \chi (\Omega_n, \theta)$.
For our choice of phonon propagator $\chi (q)$ this reduces to the Eliashberg result \cref{eq:sigma_eli}.
On the real frequency axis, the same consideration yields $\Sigma^{''} (\omega) = - g^2 \int_0^\omega \chi^{''}_L (\Omega) d \Omega$.
In the literature, $g^2 \chi^{''}_L (\Omega)$ is often denoted $\alpha^2 F(\Omega)$~\cite{marsiglio2020eliashberg,marsiglio2018eliashberg}.

\begin{figure}
    \centering
    \includegraphics[width=1\linewidth]{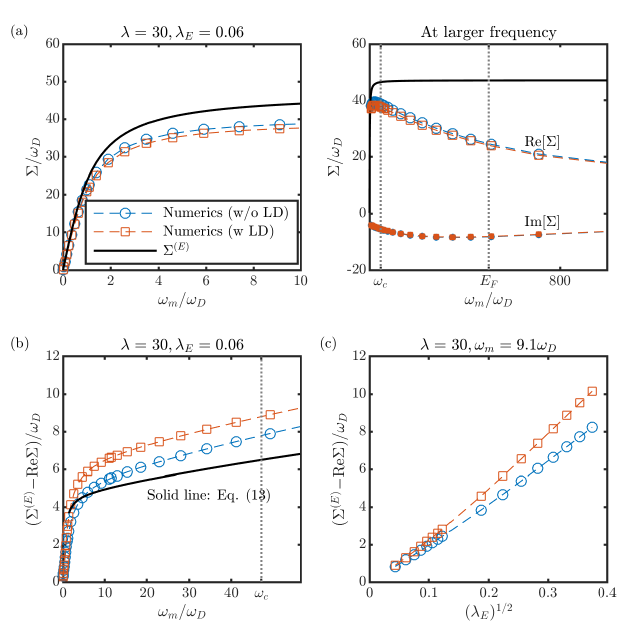}
    \caption{(Color online) One-loop electron self-energy.
     We set $E_F = 500 \omega_D$, $\lambda_E \simeq 0.06$ and $\lambda\simeq 30$.
      (a) Numerical solution of the self-consistent equation \cref{eq:sigma}
      v.s. the perturbative expression, $\Sigma^{(E)} (\omega_m)$, ~\cref{eq:sigma_eli}.
   The numerical results obtained with and without Landau damping are red and blue curves, respectively.
      The right panel in (a) displays the self-energy at larger frequencies, comparable to $E_F$.
      Both $\Re[\Sigma (\omega_m)]$ and $\Im[\Sigma (\omega_m)]$ evolve with frequency ($\Im[\Sigma (\omega_m)$ is a constant at smaller $\omega_m$ and is not shown in the left panel).
        (b) A comparison between $\Sigma^{(E)}(\omega_m) - \Sigma(\omega_m)$ and $\delta \Sigma (\omega_m)$ from \cref{eq:delta-sigma-large-frequency}.
        (c) The verification of the scaling relation $(\Sigma^{(E)}(\omega_m) - \Sigma(\omega_m))/\omega_D \propto \sqrt{\lambda_E}$ for a given $\omega_m/\omega_D\simeq 9$.
    }
    \label{fig:selfE_1loop_solution}
\end{figure}

This analysis is supported by numerical solution of the Eliashberg equations.
The numerical results are obtained by solving the self-consistent equations iteratively, where the momentum integrations are carried out without making extra approximations.
We found Fermi liquid behavior of $\Sigma (\omega_m)$ at frequencies below $\omega_D$.
The self-energy then saturates for an intermediate frequency range up to $\omega_c$, and eventually decreases at larger frequencies.
We see that the imaginary part of $\Sigma (\omega_m)$ remains weakly frequency dependent up to $\omega_m \sim E_F$. The implication is that in the whole range $|\omega_m| \leq E_F$ it can be absorbed into the renormalization of the chemical potential.
The difference between $\Re[\Sigma (\omega_m)]$ and the $\Sigma^{(E)} (\omega_m)$ is shown in Fig.~\ref{fig:selfE_1loop_solution} (b). It matches $\delta \Sigma (\omega_m)$ from Eq.~\cref{eq:delta-sigma-large-frequency}.  In particular,
we verified in \cref{fig:selfE_1loop_solution} (c) that for $\omega_m > \omega_D$, $\Re[\Sigma (\omega_m)]-\Sigma^{(E)} (\omega_m)$
 scales as $\lambda^{1/2}_{E}$, in agreement with \cref{eq:delta-sigma-large-frequency}.

\subsection{Vertex corrections}

\begin{figure}
    \centering
    \includegraphics[width=1\linewidth]{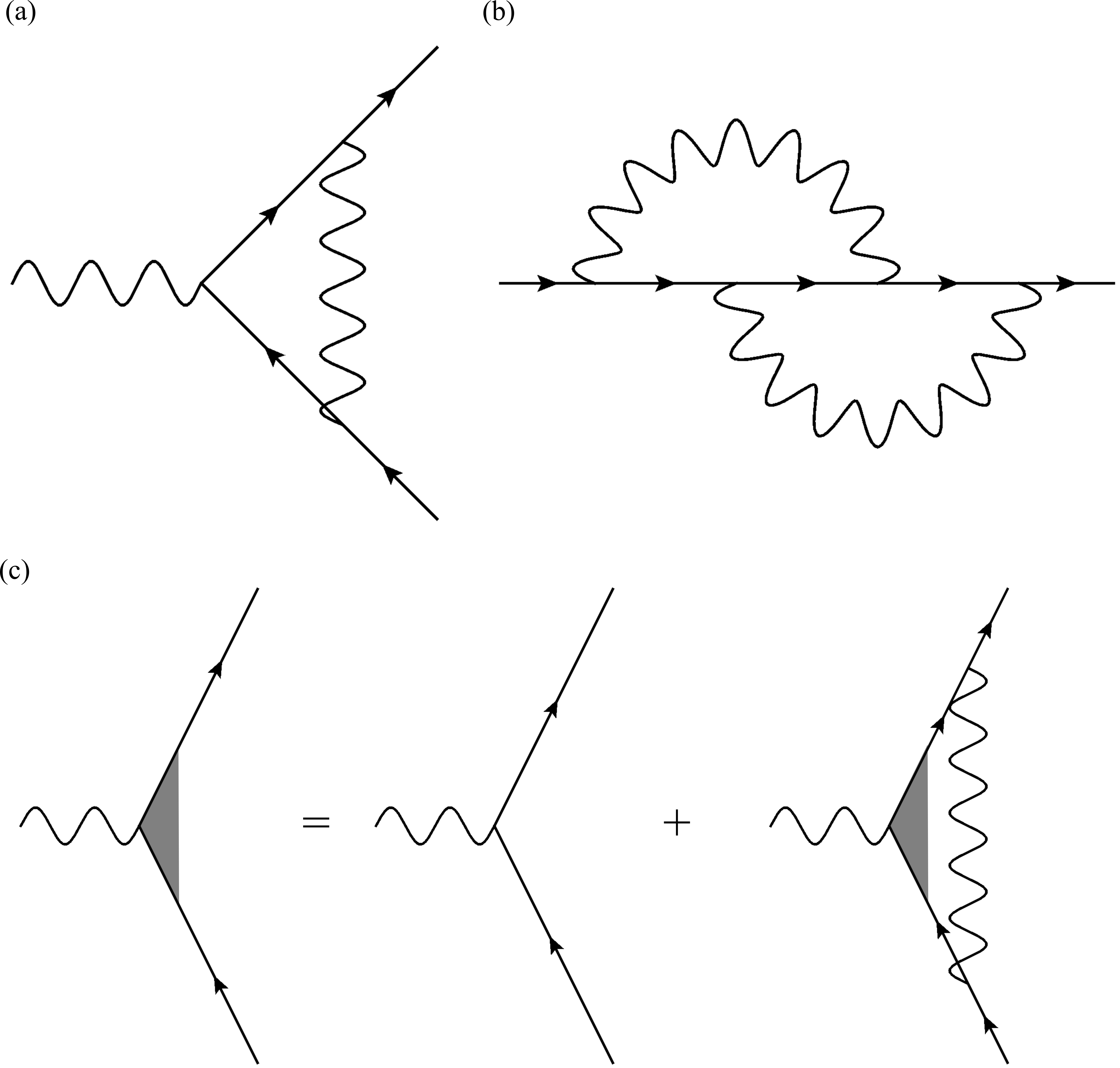}
    \caption{Relevant diagrams: (a) one-loop vertex correction, (b) two-loop electron's self-energy with vertex correction included, (c) ladder series for the vertex.}
    \label{fig:vertex_corr}
\end{figure}

The one-loop vertex correction is shown graphically in \cref{fig:vertex_corr} (a)
and the associated two-loop self-energy with vertex renormalization is shown in Fig.~\ref{fig:vertex_corr} (b). The relative strength of the two-loop self energy compared to the one-loop one determines the validity of the Eliashberg theory.
The one-loop vertex correction and the two-loop self-energy in 2D have been discussed previously in Ref.[~\onlinecite{chubukov2020eliashberg}].
Here we reproduce these earlier results and add additional details.

For external fermionic $k = (\mathbf{k}_F, \omega_m)$ and phononic $q = (\mathbf{q}, \Omega_n)$, the analytical expression for the one-loop $\Delta g/g$
is
\begin{multline} \label{eq:vertex_elph}
    \frac{\Delta g(k,q)}{g} = -i \frac{g^2}{2\pi}
    \int \frac{d \omega'_m}{(\omega_m-\omega'_m)^2 + \omega_D^2} \int \frac{d\epsilon_{k'}}{i\tilde{\Sigma} (\omega
        '_{m}
        ) - \epsilon_{k'}} \\
    \times\frac{\sgn (\omega'_m + \Omega_n)}{\sqrt{\left( v_F \left| \mathbf{q}\right| \right)^2 +\left( \tilde{\Sigma}(\omega'_m + \Omega_n) + i \left(\epsilon_{k'} + \mathbf{q}^2 / 2m \right) \right)^2}}.
\end{multline}
For $\mathbf{q}=0$, the vertex correction \emph{is not small} for any $\lambda_E$ as it
must satisfy the Ward identity associated with the conservation of the total density:
\begin{equation}
 \frac{g_\text{full}(k,q)}{g}=
    \frac{g_\text{full}(
    \omega_m,\Omega_{m},\mathbf{q}=0)}{g} = \frac{\tilde{\Sigma}(\omega_m + \Omega_n) -\tilde{\Sigma}(\omega_m)}{\Omega_n}.
    \label{eq:ward-identity}
\end{equation}
At small $\omega_m$, $\tilde{\Sigma}(\omega_m) \approx \omega_m (1+ \lambda)$, hence by the Ward identity $g_\text{full}(k,\Omega_{m},\mathbf{q}=0)/g = \lambda \gg 1$.

Putting $\mathbf{q}=0$ in \cref{eq:vertex_elph}
and setting $\Omega_n >0$ for definiteness, we perform the $\epsilon_{k'}$ integration to obtain the one-loop vertex correction
\begin{multline}
    \frac{\Delta g(
    \omega_m,\Omega_{m},\mathbf{q}=0)}{g} = g^2
    \int_{-\Omega_n}^0
    \frac{d \omega'_m}{(\omega_m-\omega'_m)^2 + \omega_D^2} \\
    \times\frac{1}{\tilde{\Sigma}(\omega'_m + \Omega_n) -\tilde{\Sigma}(\omega'_m)}.
    \label{eq:vertex_elph_1}
\end{multline}
For small $\omega_m, \Omega_n < \omega_D$, this gives $\Delta g(k,\Omega_{m},\mathbf{q}=0)/g = \lambda/(1+ \lambda)$, which approaches $1$ in the strong coupling limit $\lambda \gg 1$. Because $\Delta g(k,\Omega_{m},\mathbf{q}=0)/g \approx 1$, we need to include higher-order vertex corrections.
A simple experimentation shows that at large $\lambda$, relevant diagrams for the full vertex $g_\text{full} (k, \Omega_{m}, \mathbf{q}=0)$ form ladder series, shown in
 ~\cref{fig:vertex_corr}.
The corresponding integral equation at arbitrary $\omega_m$ and $\Omega_n$ has the form
\begin{multline} \label{eq:vertex_elph_2}
    \frac{g_\text{full}(
    \omega_m,\Omega_{m},\mathbf{q}=0)}{g} = 1 + g^2
    \int_{-\Omega_n}^0
    \frac{d \omega'_m}{(\omega_m-\omega'_m)^2 + \omega_D^2} \\
    \times	
    \frac{g_\text{full} (\omega'_m,\omega_{m},\mathbf{q}=0)/g}
    {\tilde{\Sigma}(\omega'_m + \Omega_n) -\tilde{\Sigma}(\omega'_m)}.
\end{multline}
One can straightforwardly check that the solution of this equation is given by \cref{eq:ward-identity}, in agreement with the Ward identity associated with the conservation of the total density.

For $q \sim k_F$ the result is different. Integrating over
$\epsilon_{k'}$ in \cref{eq:vertex_elph}
in infinite limits
we now obtain
\begin{widetext}
\begin{multline} \label{eq:vertex_elph_3}
    \frac{\Delta g(k,q)}{g} = \frac{g^2}{2}
    \int d \omega'_m \frac{1 -\sgn (\omega'_m + \Omega_n) \sgn (\omega'_m)}{(\omega_m-\omega'_m)^2 + \omega_D^2}
    \frac{1}{\sqrt{(v_F q)^2 + \left(\tilde{\Sigma}(\omega'_m + \Omega_n) -\tilde{\Sigma}(\omega'_m) + i q^2/(2m)\right)^2}} \\
\approx \frac{g^2}{4 E_F {\bar q}
\sqrt{1-{\bar q}^2}}
    \int d \omega'_m \frac{1 -\sgn (\omega'_m + \Omega_n) \sgn (\omega'_m)}{(\omega_m-\omega'_m)^2 + \omega_D^2}
\end{multline}
where ${\bar q} =|\mathbf{q}|/(2k_F)$. Evaluating the remaining frequency integral, we find
\begin{equation} \label{eq:vertex_elph_4}
    \frac{\Delta g(k,q)}{g} =
   \frac{\lambda_E}{4 {\bar q}}
     \frac{1}
     {\sqrt{1-{\bar q}^2}}
    \left(\arctan \left( {\omega_m+ \Omega_n \over \omega_D} \right) -
    \arctan \left( {\omega_m \over \omega_D} \right)\right).
\end{equation}
\end{widetext}
We see that for generic $\omega_m \sim \Omega_n \sim \omega_D$, $\Delta g(k,q)/g \sim \lambda_E$, as long as
 ${\bar q}$  is not too close to either $0$ or $1$.

 There is a caveat here: the vertex correction in \cref{eq:vertex_elph_4} decreases at $\omega_m, \Omega_n > \omega_D$.
On a more careful look, we found that this is an artifact of integrating over $\epsilon_{k'}$ in \cref{eq:vertex_elph}
in infinite limits instead of introducing the lower cutoff at $-E_F$. Repeating the integration with this lower cutoff we find that even at $\omega_m, \Omega_n > \omega_D$, the vertex correction remains of the form
\begin{equation}
    \label{eq:vertex_elph_4_1}
    \frac{\Delta g(k,q)}{g} = C(q) \frac{\lambda_E}{4}
   \frac{1}{{\bar q} \sqrt{1-{\bar q}^2}}, 
\end{equation}
where $C(q) = O(1)$.

\begin{figure}
    \centering
    \includegraphics[width=\linewidth]{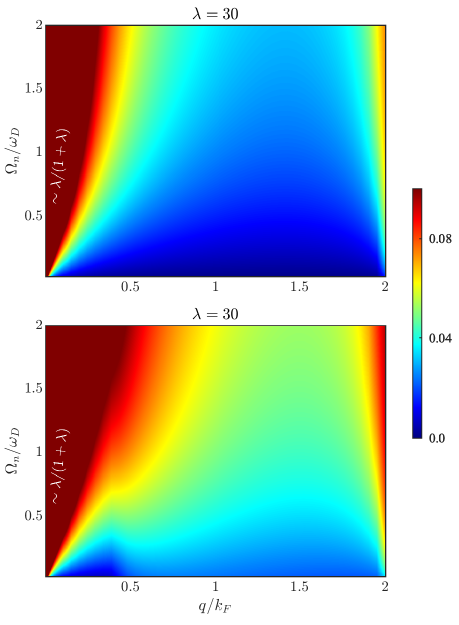}
    \caption{(Color online) Numerical results for $\Delta g/g$
        as a function of bosonic frequency $\Omega_{n}$ and momentum $q$ for
        $\lambda = 30$, $\lambda_E = 0.06$,
        and external fermionic frequency $\omega_m = 0.1 \omega_D$.
        Panels (a) and (b): the results of the calculations using infinite cutoff in the integration over dispersion,$-\infty<\epsilon_{\mathbf{k}}<\infty$, and a finite cutoff $-E_F<\epsilon_{\mathbf{k}}<\infty$, respectively.
        \label{fig:vertex_elph}}
\end{figure}

We plot the one-loop vertex correction as a function of $q$ and
$\Omega_n \sim \omega_D$
in \cref{fig:vertex_elph}.
We show the results obtained using both an infinite energy cutoff ($-\infty < \epsilon_{\mathbf{k}}<\infty$) and a finite energy cutoff ($-E_F < \epsilon_{\mathbf{k}}<\infty$).
The two results for $\Delta g/g$ do differ, most notably at $|\mathbf{q}|\sim k_F$, but both remain of order $\lambda_E$ for a generic $q$.

We next substitute $\Delta g (k,q)$ into the two-loop diagram for the self-energy,
Fig. \ref{fig:vertex_corr} (b). 
We label this contribution as  $\Sigma^{(2)}(\omega_m)$.
Analyzing the integral over phononic $q$, we find that it is singular at $q$ close to $0$ and $2k_F$,
where the vertex correction is enhanced, but the singularity is only logarithmic.
Evaluating the full integral, we find, in agreement with Ref. [~\onlinecite{chubukov2020eliashberg}],
\begin{equation}
    \Sigma^{(2)}(\omega_m) \sim \left(\lambda_E |\log{\lambda_E}|\right) \Sigma^{(E)} (\omega_m)
\end{equation}
We see that as long as $\lambda_E$ is small, the two-loop $\Sigma^{(2)}(\omega_m)$ is parametrically smaller than the Eliashberg self-energy, despite that the integral comes from $q$ where the vertex correction is singular. This singularity only accounts for $|\log{\lambda_E}|$ in the prefactor.

\begin{figure}
    \centering
    \includegraphics[width=1\linewidth]{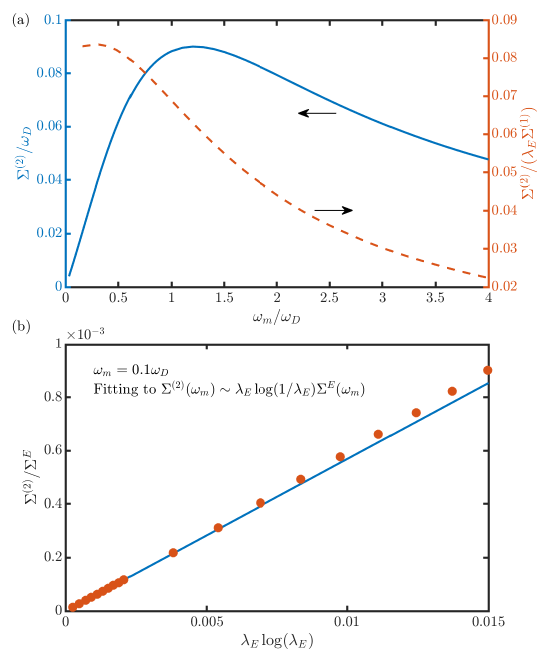}
    \caption{(a) Blue line -- the two-loop electron self-energy, obtained using infinite cutoff for integration over fermionic dispersion.
      The parameters are the same as in \cref{fig:selfE_1loop_solution}:
       $\lambda = 30$ and $\lambda_E = 0.06$.
       Brown line -- the ratio $\Sigma^{(2)}/(\lambda_E \Sigma^{(E)})$.
       The ratio approaches a constant at $\omega_m \to 0$ and evolves at
       $\omega_m \sim \omega_D$. The drop of the ratio at larger $\omega_m$ is likely an artifact of using infinite cutoff for the integration over the dispersion.
        (b) Numerical verification of the scaling of $\Sigma^{(2)}(\omega_m)/\Sigma^{(E)}(\omega_m) \sim \lambda_E |\log \lambda_E|$ at small $\omega_m$. }
    \label{fig:selfE_2loop_solution}
\end{figure}

In \cref{fig:selfE_2loop_solution} we present the result of numerical evaluation of $\Sigma^{(2)}(\omega_m)$ using an infinite cutoff in the integration over $\epsilon_{k'}$. We see that the ratio $\Sigma^{(2)}/\Sigma^{(E)}$ is approximately constant at $\omega_{m} \leq \omega_D$ and as a function of $\lambda_E$ does scale as $\lambda_E |\log{\lambda_E}|$. The drop of the ratio $\Sigma^{(2)}/\Sigma^{(E)}$ at larger frequencies is likely an artifact
of an infinite cutoff. In any case, the ratio $\Sigma^{(2)}/\Sigma^{(E)}$ is small at small $\lambda_E$.
 For completeness we computed this ratio at $\lambda_E = 1$ and found that it still remains numerically small over the wide range of  $\omega_D < \omega_m < E_F$ ($\Sigma^{(2)}/\Sigma^{(E)}\simeq 0.062$ at $\omega_{m}=\omega_{D}$).

For comparison with the Ising-nematic/Ising-ferromagnetic case below, we analyze how $\Delta g/g$ depends on characteristic momenta and frequencies.
For definiteness we set $\mathbf{k} = \mathbf{k}_F$, $\Omega_n \sim \omega_D$ and vary $\omega_m$ and  $q_{\perp}$ and $q_{\parallel}$, which are typical momentum components transverse and along the Fermi surface.  At $\omega_D < \omega_m < \omega_c =(\pi/2) \lambda_E E_F $, where
the self-energy $\Sigma (\omega_m)$ exceeds bare $\omega_m$,
 which are typical momentum components transverse and along the Fermi surface. After a simple experimentation
 we found after a simple experimentation
\begin{equation} \label{eq:vertex_elph_5}
    \frac{\Delta g(k,q)}{g} \sim \frac{\omega_c}
    {\omega_c + v_F q_{\perp} + \frac{q^2_{\parallel}}{2m}}.
\end{equation}
Typical $v_F q_{\perp}$ are of order $\omega_c$, typical $q^2_{\parallel}/(2m)$ are of order $E_F$. For these momenta,
\begin{equation} \label{eq:vertex_elph_6}
    \frac{\Delta g(k,q)}{g} \sim \frac{v_F q^{\text{typ}}_{\perp}}{v_F q^{\text{typ}}_{\perp} + \frac{(q^{\text{typ}}_{\parallel})^2}{2m}}.
\end{equation}
The ``fast electrons/slow bosons'' criterion requires typical $q_{\perp}$ to be much smaller than typical $q_{\parallel}$. This holds for $\lambda_E \ll 1$ because typical $q^{\text{typ}}_{\perp} \sim \lambda_E k_F$, while $q^{\text{typ}}_{\parallel} \sim k_F$. The strength of the vertex correction in \cref{eq:vertex_elph_6} is, however, determined by the ratio of typical energies $v_F q^{\text{typ}}_{\perp}/((q^{\text{typ}}_{\parallel})^2/2m)$ rather than
typical momenta. In general, the ratios of typical energies and typical momenta are not the same, but in the electron-phonon case,
  $v_F q^{\text{typ}}_{\perp}/((q^{\text{typ}}_{\parallel})^2/2m) \sim \lambda_E$ is the same as $q^{\text{typ}}_{\perp}/q^{\text{typ}}_{\parallel}$ because $q^{\text{typ}}_{\parallel} \sim k_F$.
As a consequence, the single parameter $\lambda_E$ allows one to simplify the Eliashberg equations and
keep
vertex corrections small.

For larger $\omega_m$, we found
\begin{equation} \label{eq:vertex_elph_5_1}
    \frac{\Delta g(k,q)}{g} \sim \frac{\omega_c}
    {\omega_m + v_F q^{\text{typ}}_{\perp} + \frac{(q^{\text{typ}}_{\parallel})^2}{2m}}.
\end{equation}
Now typical $v_F q_{\perp}$ are of order $\omega_m$, while typical $q^2_{\parallel}/(2m)$ are still of order $E_F$. For these momenta,
\begin{equation} \label{eq:vertex_elph_6_1}
    \frac{\Delta g(k,q)}{g} \sim \frac{\omega_c}{\omega_m + E_F} \sim \lambda_E \frac{1}{1+ \frac{\omega_m}{E_F}}
    \end{equation}
We see that the vertex correction remains of order $\lambda_E$.

\subsection{Summary of \cref{sec:el-ph}}

There are three energy scales in the electron-phonon  problem: the bosonic energy $\omega_D$, the dimension-full electron-phonon coupling $g$ and the Fermi energy $E_F$.
This allows one to introduce two dimensionless ratios $\lambda = g^2/\omega^2_D$ and $\lambda_E = g^2/(E_F \omega_D) = \lambda \omega_D/E_F$. The latter is a small parameter for the Eliasberg theory.
The strong coupling regime occurs at $\lambda \gg 1$, $\lambda_E  \ll 1$.
In this regime, the system displays Fermi liquid behavior at $\omega < \omega_D$, non-Fermi liquid behavior with $\Sigma (\omega_m)  \approx (\pi/2) \lambda \omega_D = (\pi/2) \lambda_E E_F$ at $\omega_D < \omega_m < \omega_c$, where
$\omega_c \sim \lambda \omega_D \sim E_F \lambda_E$, and Fermi-gas behavior at larger frequencies.

The three key results for electron-phonon system, which we will later use in comparison with the behavior near a nematic QCP, are the following.

First, Eliashberg theory is rigorously justified even at strong coupling $\lambda \gg 1$, as long as the Eliashberg parameter $\lambda_E \ll 1$.
Two-loop vertex corrections to self-consistent one-loop Eliashberg theory are small in $\lambda_E$ for all frequencies $\omega_m$.

Second, the same small parameter $\lambda_E$ also simplifies the calculations within Eliashberg theory:
the Landau damping of phonons can be neglected in the calculation of the self-energy, and
the self-energy $\Sigma(\mathbf{k}, \omega_m)$ can be approximated by the local $\Sigma (\omega_m)$ and computed perturbatively rather than self-consistently.
 This holds for $\omega_m < E_F$. At larger $\omega_m$,
 ``fast electron/slow boson''
 criterion is not valid, and the Eliashberg equation (\ref{eq:sigma}) has to be re-analyzed.

Third, vertex corrections to Eliashberg theory become $O(1)$ at $\lambda_E = O(1)$, and become parametrically large at $\lambda_E >1$,
except for the smallest frequencies $\omega_m <\omega_D/\lambda_E$, where they remain small.
Because $\lambda_E = g^2/(\omega_D E_F)$, this result implies that one cannot extend the Eliashberg theory of electron-phonon interaction to $\omega_D =0$.
For a finite $E_F$, Eliashberg theory is valid only when $\omega_D$ exceeds a certain value.
At smaller $\omega_D$, a
new description is required.

\begin{figure}
    \centering
    \includegraphics[width=0.9\linewidth]{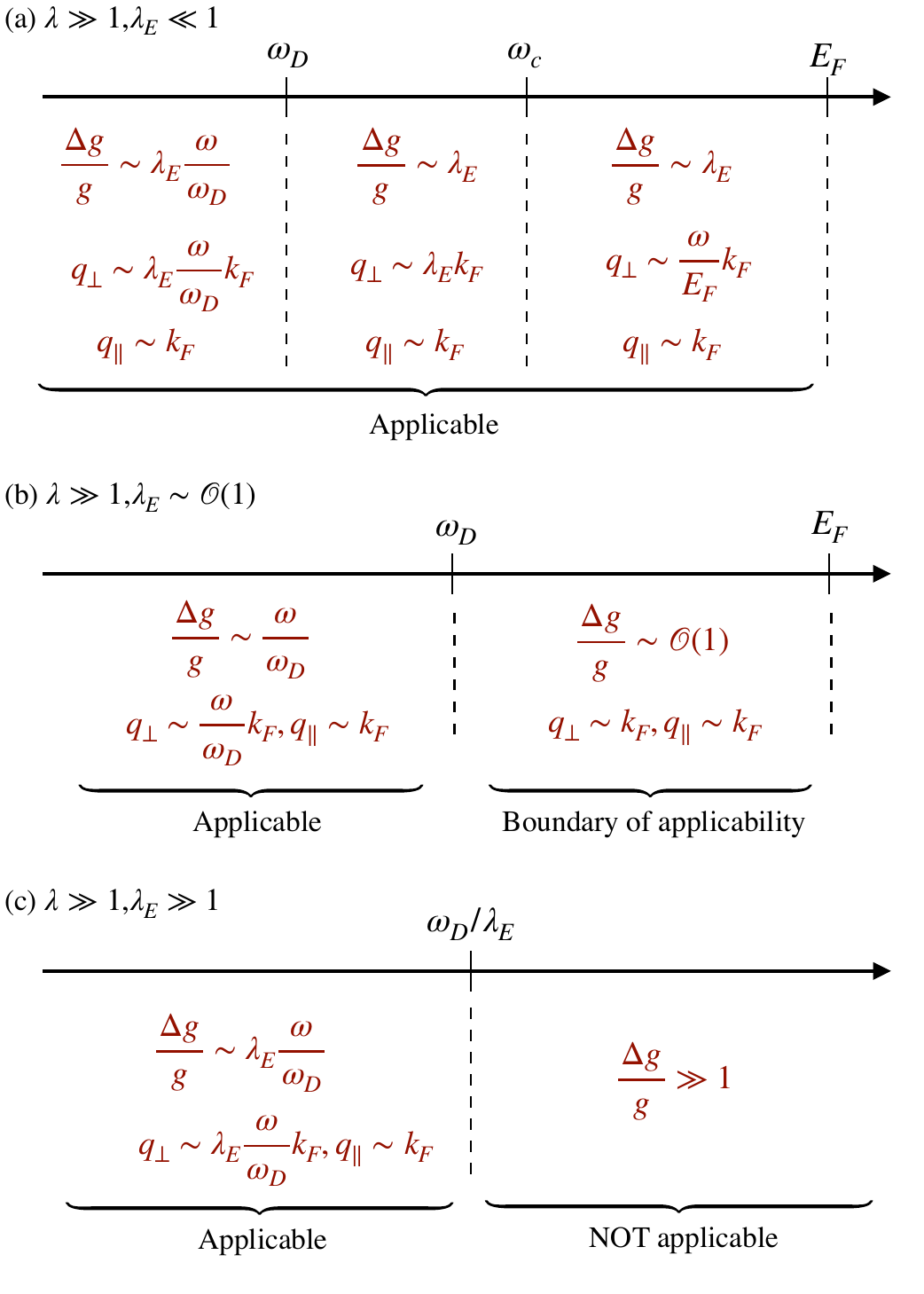}
    \caption{Illustration of the applicability of Eliashberg theory
        for electron-phonon interactions
        at different energy scales: (a) $\lambda \gg 1$, $\lambda_E \ll 1$; (b) $\lambda \gg 1$, $\lambda_E \sim 1$; (c) $\lambda \gg 1$, $\lambda_E \gg 1$.}
    \label{fig:pd_elph}
\end{figure}

We summarize these results in \cref{fig:pd_elph}.

\section{Quantum critical metal}
\label{sec:qcm}

We now analyze the validity of an Eliashberg-type description of a system of electrons interacting with soft collective bosons representing order parameter fluctuations.
For definiteness, we focus on the system near a
${\bm q}=0$
Ising-ferromagnetic or Ising-nematic instability (see e.g., Ref.~\onlinecite{Zhang2023} and references therein).
The two cases differ by the form-factor for fermion-boson coupling (it has d-wave form in the Ising-nematic case).
We verified that the form-factor does not play a critical role in our considerations and can be safely set to one
(see also Ref. \cite{Klein_2020_a}).
 We do not consider $SU(2)$ ferromagnetic case as the ordering transition there is affected by non-analytic corrections to spin susceptibility~\cite{Chubukov2009}.

The set of Eliashberg-type equations is the same as \cref{eq:sigma,eq:pi}.
Like for the electron-phonon case, it can be either derived diagrammatically~\cite{acs,*finger_2001,Klein_2018_a} as self-consistent one-loop equations, or obtained as stationary solutions of the Luttinger-Ward functional~\cite{Zhang2023}.
The bare fermionic Green's function is the same as in \cref{eq:bare-gfs}.
For the bare bosonic propagator we choose the conventional Ornstein-Zernike form
\begin{equation}
    \chi_0 (\mathbf{q}) = \frac{
        \chi_0}{\xi^{-2} + |\mathbf{q}|^2}.
    \label{eq:a2_1}
\end{equation}
We will define a related frequency scale $\omega^{*}_{D}=v_{F}\xi^{-1}$.
The full generic $\chi_0 (\mathbf{q})$ also contains a dynamical $(\Omega_{n}/v_F)^2$ term,
however, this term only becomes
 relevant at frequencies above the upper limit of quantum-critical behavior (see below).
For this reason, we neglect the $\Omega^2_n$ term in $\chi_0$ in our analysis.

We label the fermion-boson coupling as $\tilde{g}$ and introduce the effective dimension-full interaction $g^* = \tilde{g}^2 \chi_0$.
Like in the electron-phonon case, there are three energy scales in this model:
the effective coupling $g^*$, the bosonic energy $\omega^*_D$, and the Fermi energy $E_F$.
We again assume that $E_F$ is the largest scale in the problem, i.e., set $E_F \gg \omega^*_D, g^*$.
 Like before, one can construct two dimensionless ratios out of these couplings. We choose them to be
\begin{equation}
    \lambda^* = \frac{g^*}{4\pi \omega^*_D},\quad \lambda^*_E = \frac{g^*}{E_F}.
    \label{eq:aa1}
\end{equation}
By construction, $\lambda^*_E \ll 1$, but $\lambda^*$ can be either small or large.
We will see that the self-energy at the lowest $\omega_m$ is $\Sigma = \lambda^* \omega_m$, i.e., $\lambda^* \sim 1$ separates
 weak coupling ($\lambda^* <1$) and strong coupling ($\lambda^* >1$) regimes, respectively.

\subsection{Solution of Eliashberg equations}

The analysis of the Eliashberg equations for fermions near a $q=0$ QCP has been done before in various contexts~\cite{PALee1989,*Monien1993,*Nayak1994,*Altshuler1994,*Kim_1994,rech_2006,DellAnna2006,*Metzner2003,Maslov2010,Fradkin2016,Esterlis2021, 10.21468/SciPostPhys.13.5.102,Klein2018,*Klein2019,*Klein2020,*Klein2022,paper_1,Zhang2023}.
We list the existing results and present some new ones that will allow a direct comparison with the electron-phonon case.

We first discuss the calculation of the fermionic self-energy and bosonic polarization.
Like for the electron-phonon case, we assume and then verify that the self-energy $\Sigma (k, \omega)$ can be approximated by a local $\Sigma (\omega_m)$.
For such a self-energy, earlier calculations found that for $v_F |\mathbf{q}| \gg |\Omega_n|$, and both much smaller than $E_F$,
the polarization can be written as a sum of static and dynamic terms
$\Pi (\mathbf{q}, \Omega_n) = \Pi (\mathbf{q}, 0) + \delta \Pi (\mathbf{q}, \Omega_n)$, where $\delta \Pi (\mathbf{q}, \Omega_n)$ has the form of Landau damping:
\begin{equation}
    \delta
    \Pi (\mathbf{q}, \Omega_n) = \frac{
        g^* k_F}{\chi_0
        \pi v^2_F} \frac{|\Omega_n|}{|\mathbf{q}|}.
    \label{eq:delta-pi-ising}
\end{equation}
This is similar to the
electron-phonon polarization,
\cref{eq:delta-pi} at $q \ll k_F$.
Like there, the prefactor for the Landau damping term does not depend on the fermionic self-energy, i.e., \cref{eq:delta-pi-ising} has the same form for free fermions and for interacting fermions.
We incorporate the static $\Pi (\mathbf{q}, 0)$ into $\chi_0 (|\mathbf{q}|)$, like we did for the electron-phonon case.
Substituting into $\chi (q)$, we then obtain
\begin{equation}
    \chi(q) = \frac{\chi_{0}}{\xi^{-2} + |\mathbf{q}|^{2} + \alpha\frac{|\Omega_n|}{|\mathbf{q}|}},
    \label{eq:chi-ising-renormalized}
\end{equation}
where $\alpha = \frac{g^{*}k_{F}}{\pi v_{F}^{2}}$.
Because relevant $\mathbf{q}$ are much smaller than $k_F$, the
dynamical Landau damping term in $\chi(q)$ becomes relevant when $\Omega_n$ exceeds the scale
\begin{equation}
    \omega^* =
    \frac{(\omega^*_D)^3}{g^* E_F} \sim \frac{\omega^*_c}{(\lambda^*)^3},
    \label{eq:aa4}
\end{equation}
where $\omega^*_c \sim E_F (\lambda^*_E)^2$ (the exact definition with the prefactor is in \cref{eq:omega-0} below).
Because we consider
$\lambda_E^*$
to be small, we have $\omega^*_c \ll E_F$.
We will see that at strong coupling, when
$\lambda^* >1$, $\omega^*$ is the upper edge of Fermi liquid behavior, while $\omega^*_c > \omega^*$ is the upper edge of quantum-critical non-Fermi liquid behavior.
This identification implies that the Landau damping term in \cref{eq:chi-ising-renormalized} is essential outside of the Fermi liquid regime.
This distinguishes this case from the electron-phonon one, where relevant $q \sim k_F$ and the Landau damping term is irrelevant for all frequencies, as long as $\lambda_E$ is small.

We now substitute $\chi (q)$ from \cref{eq:chi-ising-renormalized} into the formula for the self-energy, \cref{eq:sigma}.
As before, we assume and verify that $\Sigma (\mathbf{k}, \omega_m) \approx \Sigma (\mathbf{k}_F, \omega_m) = \Sigma(\omega_m)$.
The self-consistent equation for $\Sigma(\omega_m)$ is
\begin{widetext}
\begin{equation}
    \Sigma (\omega_m) = g^*\int \frac{d \Omega_n}{2\pi} \int \frac{d^2 \mathbf{q}}{4\pi^2}
    \frac{1}{\xi^{-2} +|\mathbf{q}|^2 + \alpha\frac{|\Omega_n|}{|\mathbf{q}|}} \frac{1}{i\tilde{\Sigma}(\omega_m + \Omega_n) -\epsilon_{\mathbf{k}_F + \mathbf{q}}},
    \label{eq:ising-self-energy-eqn}
\end{equation}
\end{widetext}
where, as before, $\tilde{\Sigma} (\omega_m) = \omega_m + \Sigma (\omega_m)$.
A straightforward analysis of the relevant scales in this equation shows that relevant $\Omega_n \sim \omega_m$, relevant $\epsilon_{\mathbf{k}_F + \mathbf{q}} \sim \tilde{\Sigma} (\omega_m)$, and relevant $|\mathbf{q}| \sim \xi^{-1}$ for $\omega_m < \omega^*$ and $|\mathbf{q}| \sim (\alpha |\omega_m|)^{1/3}$ for $\omega_m > \omega^*$.
For the fermionic dispersion $\epsilon_{\mathbf{k}_F + \mathbf{q}}$ we will be using
\begin{equation}
    \epsilon_{\mathbf{k}_F+\mathbf{q}} = v_{F}q_{\perp} + \frac{q_{\parallel}^{2}}{2m},\label{eq:dispersion}
\end{equation}
where $q_{\perp}$ and $q_{\parallel}$ are momentum components perpendicular and parallel to the Fermi surface (along $\mathbf{k}_F$ and perpendicular to $\mathbf{k}_F$, respectively).
Comparing relevant scales, we immediately see that for $\lambda^*_E \ll 1$, relevant
 $q_{\parallel}$ are much larger than $q_{\perp}$ as long as
$\omega_m < \omega_{\text{max}} \equiv
    (g^*E_F)^{1/2} \sim
    \omega^*_c /(\lambda^*_E)^{3/2}$.
For frequencies below $\omega_{\text{max}}$ we then can factorize the momentum integration, i.e.,
re-express $d^2 \mathbf{q}$ as $(1/v_F)d \epsilon_{\mathbf{k}_F + \mathbf{q}} d q_{\parallel}$ and integrate the fermionic propagator over $\epsilon_{\mathbf{k}_F + \mathbf{q}}$ and bosonic $\chi (|\mathbf{q}|, \Omega_n) \approx \chi (|q_{\parallel}|, \Omega_n)$ over $q_{\parallel}$.
Using
\begin{equation}
    \int \frac{d \epsilon}{i \tilde{\Sigma} (\omega_m + \Omega_n) - \epsilon} = -i \pi {\sgn} (\omega_m + \Omega_n)
\end{equation}
we find that the right-hand side of \cref{eq:sigma} does not depend on the self-energy, i.e., has the same form as for free fermions.
Completing the integration, we obtain
\begin{equation}
    \Sigma^{(E)} (\omega_m) = \lambda^* \omega_m F\left(\frac{\omega_m}{\omega^*}\right)
    \label{eq:sigma-eliashberg-ising}
\end{equation}
in terms of a function $F$, where $F(0) =1$ and $F(x \gg 1) =
    ((4\pi)^{1/3}/\sqrt{3}) /x^{1/3}$.
One can obtain the full analytical formula for $F(x)$, but it is rather cumbersome and not particularly enlightening so we do not present it here.

\begin{figure}
    \centering
    \includegraphics[width=0.9\linewidth]{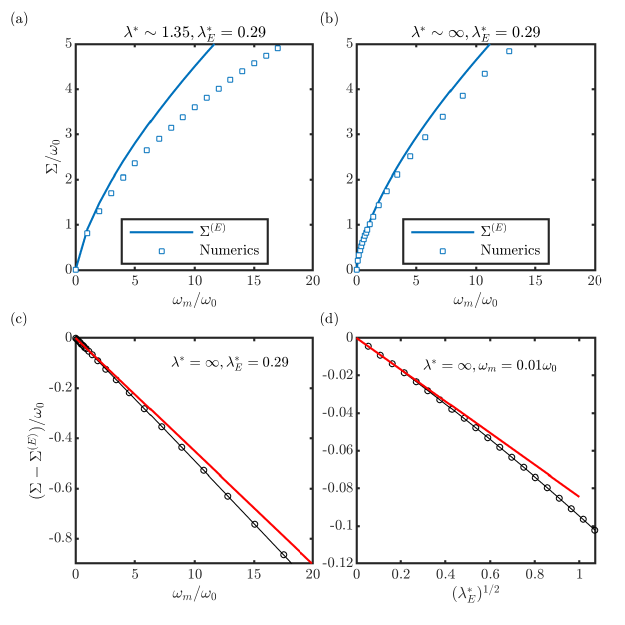}
    \caption{
      (Color online)
      Numerical results for the self-energy in Eliashberg theory
      for $\lambda^{*}_{E}=0.29$ at (a) intermediate coupling $\lambda^{*} = 1.35$ and (b) infinitely strong coupling $\lambda^* =\infty$ at a QCP.
        The solid line is  the perturbative expression $\Sigma^{(E)}$, Eq. (\ref{eq:sigma-eliashberg-ising}), obtained by
        factorizing the momentum integration, and squares are the numerical solutions of the full equation (\ref{eq:ising-self-energy-eqn}) for $\Sigma (\omega_m)$.
        (c) The difference between  $\Sigma(\omega_m)$ and $\Sigma^{(E)}(\omega_m)$ at the  QCP ($\lambda^* = \infty$). Red solid line is the analytic result,  \cref{eq:delta-sigma-ising}. (d) Verification of the scaling relation $\Sigma(\omega_m) - \Sigma^{(E)}(\omega_m) \propto (\lambda_E^*)^{1/2}$ (red line) at a particular $\omega_m$.
    \label{fig:1lp_in}}
\end{figure}
At small frequencies, $\Sigma^{(E)} (\omega)$ has the Fermi liquid form,
$\Sigma (\omega_m) = \lambda^* \omega_m$.
For large
$\lambda^*$,
$\Sigma^{(E)} (\omega_m) \gg \omega_m$.
In this regime, relevant
$q_{\parallel} \sim \xi^{-1}$ while relevant
$q_{\perp} \sim (\omega_{m}/\omega^*)(\omega_D^*/E_F) \xi^{-1}$
are far smaller.
This is consistent with the ``fast electrons/slow bosons'' criterion.
In this regime, we also have typical
$q^2_{\parallel}/k_F \sim (\omega_D^*/E_F)\xi^{-1}$.
This will be relevant to our analysis of vertex corrections below.

At larger $\omega _{m}> \omega^*$, the self-energy crosses over to a non-Fermi liquid, quantum-critical form
$\Sigma^{(E)} (\omega_m) = \omega^{2/3}_m (\omega^*_c)^{1/3}$, where
\begin{equation}
        \omega^*_c =\frac{1}{16\pi^
            2
            \sqrt{27}} \frac{(g^*)^2}{E_F} = \frac{1}{
            16\pi^2
            \sqrt{27}}E_F (\lambda^*_E)^2.
 \label{eq:omega-0}
\end{equation}
In terms of $\omega^*_c$,
\begin{equation}
\omega^* = \frac{\sqrt{27}}{8} \frac{\omega^*_c}{(\lambda^*)^3}, \omega_{max} = 16\pi^2 \sqrt{27} \frac{\omega^*_c}{(\lambda^*_E)^{3/2}}
    \end{equation}
In the quantum-critical regime, relevant $q_{\parallel} \sim (\alpha \omega_m)^{1/3}$ and relevant $q_{\perp} \sim \omega^{2/3}_m \omega^{1/3}_0/v_F$.
These can be re-expressed as $q_{\parallel} \sim k_F (\omega_m/\omega^*_c)^{1/3} \lambda^*_E$ and $q_{\perp} \sim k_F (\omega_m/\omega^*_c)^{2/3} (\lambda^*_E)^2$.
We see that relevant $q_{\parallel}$ are again larger than $q_{\perp}$, i.e., the ``fast electrons/slow bosons'' criterion is well satisfied when $\lambda^*_E$ is small.
The conditions look similar to the Fermi liquid regime, but there is one important difference: relevant values of $q^2_{\parallel}/k_F$ are now comparable to relevant $q_{\perp}$.
We will discuss below how this affects vertex corrections.

At even larger $\omega_m > \omega^*_c$, the self-energy becomes smaller than $\omega_m$, although it still preserves its $\omega^{2/3}_m$ form.
In this last regime relevant $q_{\parallel} \sim (\alpha \omega_m)^{1/3}$, like before, but relevant
 $q_{\perp} \sim \omega_m / v_F$.
These can be re-expressed as $q_{\parallel} \sim k_F (\lambda^*_E)^{1/2} (\omega_m/\omega_\text{max})^{1/3} $ and $q_{\perp} \sim k_F (\lambda^*_E)^{1/2} (\omega_m/\omega_\text{max})$.
The ``fast electrons/slow bosons'' criterion is satisfied as long as $\omega_m < \omega_\text{max}$.

At $\omega_m > \omega_\text{max}$ the criterion of fast electrons and slow bosons is no longer valid and, as a result, one cannot factorize the momentum integration.
In this regime, however, the $\Omega^2_m$ term in the bare $\chi_0 (q)$ cannot be neglected.
We don't analyze this high-frequency regime here.

\subsection{Subleading terms in the self-energy}
\label{sec:subleading_in}

As for the electron-phonon case,
 there are three
corrections to the
 perturbative one-loop self-energy.
 They
come
  from (i) the momentum dependence of $\Sigma (\mathbf{k}, 0)$, (ii) from
  non-factorization of momentum integration and (iii)  from
 keeping $\tilde{\Sigma} (\omega_m + \Omega_n)$ in the right-hand side of \cref{eq:ising-self-energy-eqn}.
 The momentum dependent part of the self-energy is $\Sigma (\mathbf{k}, 0) \sim \epsilon_k (\lambda^*_E)^{1/2}$ ~\cite{rech_2006}.
It gives a small renormalization of the dispersion at $\lambda^*_E \ll 1$.
The correction
  from non-
 factorization of momentum integration
  is of order of the ratio of relevant scales $(q_{\perp}/q_{\parallel})^2$.
In the quantum-critical regime
this yields a relative correction to the self-energy of
order $(\omega/\omega^*_c)^{2/3} (\lambda^*_E)^2$, which is at most
$O((\lambda^*_E)^2)$.
In the Fermi liquid regime the correction is even smaller.
In the Fermi gas regime the correction increases and becomes of order one at $\omega \sim \omega_{\text{max}}$, which, we remind, is the largest frequency up to which the concept the ``fast electron/slow boson'' is applicable.

 The correction $\delta \Sigma (\omega_{m})$ from keeping $\tilde{\Sigma} (\omega_m + \Omega_n)$
  is
  of order
  $\omega_m (\lambda_E)^{1/2}$ for all $\omega_m < \omega_{{\text{max}}}$.
We verified that it comes from
  internal $\Omega_n \sim \omega_{\text{max}}$ and momenta
 $q_{\perp} \sim q_{\parallel} \sim k_F (\lambda_E^*)^{1/2}$.
For such $\Omega_n$, $\tilde{\Sigma} (\omega_m + \Omega_n) \approx \omega_m + \Omega_n$, i.e., the actual self-energy of an internal fermion is irrelevant.
In explicit form we find for $\delta \Sigma (\omega_{m})$
\begin{equation}
    \delta \Sigma(\omega_m)
    = - \omega_m (\lambda_E)^{1/2} J
    \label{eq:delta-sigma-ising}
\end{equation}
where
\begin{equation}
    J = \frac{1}{2\pi^2} \int_0^\infty dx \int_0^\infty dy \frac{x^2 y}{(4x^2 +y^2)^{3/2} \left(x^3 + \frac{y}{4\pi}\right)} \approx 0.084.
\end{equation}
 and the variables are  $y = \Omega_n/\omega_{\text{max}}$ and
 $x = q/(k_F (\lambda_E^*)^{1/2})$.
We caution, however, that at $y = O(1)$, i.e., at  $\Omega_n \sim \omega_{\text{max}}$, the $(\Omega_{m}/v_F^2)$ term in $\chi (q)$, which we neglected, becomes comparable with the Landau damping, and this may change the value of $J$.
  Still,
it remains a number of order one, and $\delta \Sigma(\omega_m)$ is much smaller than $\Sigma^{(E)} (\omega_m)$.

In \cref{fig:1lp_in}, we show the numerical result of the self-energy for a small but finite $\lambda_E^*$. The solid line is $\Sigma^{(E)} (\omega_m)$ at $k=k_F$ and the dotted line is
 the full self-consistent numerical result
$\Sigma (\omega_m) = \Sigma^{(E)} (\omega_m) + \delta \Sigma (\omega_m)$. We see that $\delta \Sigma (\omega_m)$ is linear in $\omega_m$,
and the prefactor scales as $(\lambda_E^*)^{1/2}$,
as in (\ref{eq:delta-sigma-ising}).
The numerical data matches \cref{eq:delta-sigma-ising} very well.

\subsection{Vertex correction}

Vertex corrections to Eliashberg theory have been discussed in detail in Ref.~\onlinecite{rech_2006}.
We present several new results below.

The lowest order vertex correction is given by the diagram in
\cref{fig:vertex_corr} (a).
In explicit form
\begin{widetext}
\begin{equation} \label{eq:vertex_qcp}
    \frac{\Delta g(k,q)}{g} = g^{*}\int \frac{d\Omega'_m d^{2}\mathbf{q}'}{(2\pi)^{3}}
    \frac{1}{i\tilde{\Sigma}(\omega_m+\Omega'_n)-\epsilon_{\mathbf{k}+\mathbf{q}'}}
    \frac{1}{i\tilde{\Sigma}(\omega_m+\Omega'_n+\Omega_n)-\epsilon_{\mathbf{k}+\mathbf{q}'+\mathbf{q}}}
    \frac{1}{\left|\mathbf{q}'\right|^{2}+\alpha\frac{\left|\Omega'_n\right|}{\left|\mathbf{q}'\right|}},
\end{equation}
\end{widetext}
where we recall, $q = (\mathbf{q}, \Omega_n)$ is the external bosonic momentum and $k=(\mathbf{k}, \omega_m)$ is the external fermionic momentum.

Like for the electron-phonon case, vertex correction at $\mathbf{q}=0$ has to obey the Ward identity, \cref{eq:ward-identity}, associated with the conservation of the particle number. We show that this indeed holds.

In the Fermi liquid regime at $\omega_m < \omega^*$, an explicit calculation of $\Delta g (k,\Omega_{n}, \mathbf{q}=0)/g$ yields,
\begin{equation}
    \frac{\Delta g(k,\Omega_{n},\mathbf{q}=0)}{g} = \frac{\lambda}{1 + \lambda}
\end{equation}
At large $\lambda$, the one-loop vertex correction is close to one.
To get the full vertex $g_\text{full} (k,\Omega_{m},\mathbf{q}=0)$ one has to sum the series of ladder vertex correction diagrams, just as we did for the electron-phonon case.
The summation yields the expected result
\begin{equation}
    \frac{g
            _{\text{full}}
        (k,\Omega_{n}, \mathbf{q}=0)}{g} = \frac{1}{1-\frac{\lambda}{1 + \lambda}}= 1 + \lambda,
    \label{eq:g_full_fl_regime}
\end{equation}
 which is equal to $(\tilde{\Sigma} (\omega_m + \Omega_n) - \tilde{\Sigma} (\omega_m))/\Omega_n$ in the FL frequency regime,
and thus satisfies the Ward identity, \cref{eq:ward-identity}.

In the quantum-critical regime, the one-loop vertex correction is again of $O(1)$, but has a more complicated structure.
Introducing dimensionless variables $x = \omega_m/\Omega_n$ and
$x' = - (\omega_m + \Omega'_n)/\Omega_n$ and setting both $\omega_m$ and $\Omega_n$ to be positive, we find
 one-loop vertex correction in the form
\begin{multline} \label{eq:ladder}
  \frac{\Delta g (x,\Omega_n)}{g} = \frac{2}{3}\int_{0}^{1}\frac{dx'}{\left|x^\prime-x \right|^{1/3}}\\
  \times \frac{1}{(1-x^\prime)^{2/3}+(x^\prime)^{2/3}+ \left(\frac{\Omega_n}{\omega^*_c}\right)^{1/3}}.
\end{multline}
The ladder series of vertex corrections in this situation
can be re-expressed as an integral equation for
$g_\text{full} (x, 
\Omega_{n})$:
\begin{widetext}
\begin{equation}
    \label{eq:ladder_1}
    g_\text{full} 
    (x, \Omega_{n})
    = 1 +\frac{2}{3}\int_{0}^{1}\frac{d x^\prime} {\left|
    x^\prime-x \right|^{1/3}} \frac{g_\text{full} (x^\prime, \Omega_n)}{(1-x^\prime)^{2/3}+(x^\prime)^{2/3}+
        \left(\frac{\Omega_n}{\omega^*_c}\right)^{1/3}}
\end{equation}
This is an integral equation in the variable $x$, while $\Omega_n$ just acts as a parameter.
One can easily verify that the solution of \cref{eq:ladder_1} is
\begin{equation}
    g_\text{full} (x, \Omega_n) = \left(\frac{\omega^*_c}{\Omega_n}\right)^{1/3}
    \left((1-x)^{2/3}+x^{2/3}+ \left(\frac{\Omega_n}{\omega^*_c}\right)^{1/3}\right).
    \label{eq:gfull-x}
\end{equation}
In the original variables \cref{eq:gfull-x} becomes
\begin{equation}
    g_\text{full} (\omega_m, \Omega_n) = 1 + \frac{\left((\omega_m+ \Omega_n)^{2/3} - (\omega_m)^{2/3}\right) \omega^{2/3}_0}{\Omega_n} = \frac{\tilde{\Sigma} (\omega_m + \Omega_n) - \tilde{\Sigma} (\omega_m)}{\Omega_n}
    \label{eq:gfull-omega}
\end{equation}
\end{widetext}
as in \cref{eq:ward-identity}.

For finite $\mathbf{q}$, the result is more involved.
 The  one-loop vertex correction for $\mathbf{k}$ on the Fermi surface is a scaling function of five variables
\begin{equation}
    \frac{\Delta g(k,q)}{g} =
    \Phi\left(\frac{\omega_m}{\Omega_n},\frac{v_{F}q_{\perp}}{\tilde{\Sigma}(\omega_m)},
    \frac{ q_{\parallel}^{2}/2m}{\tilde{\Sigma}(\omega_m)}, \frac{\omega_m}{{\tilde \Sigma} (\omega_m)}, \frac{(\omega^*_D)^2}{E_F {\tilde \Sigma}(\omega_m)}\right)
    \label{eq:delta-g-scaling}
\end{equation}

In the Fermi liquid regime at $\omega_m < \omega^*$,
 typical $v_{F} q_{\perp} \sim \Sigma (\omega_m) = \lambda^* \omega_m$ 
and
typical $(q^2_{\parallel}/2m)/\Sigma (\omega_m) \sim \omega^*/\omega_{m}$.
 We verified numerically
that
in this case
$\Phi$ is
of order $\omega_{m}/\omega^*\ll 1$, i.e., $\Delta g(k,q)/g \sim \omega_{m}/\omega^* \ll 1$.

\begin{figure}
    \centering
    \includegraphics[width=\linewidth]{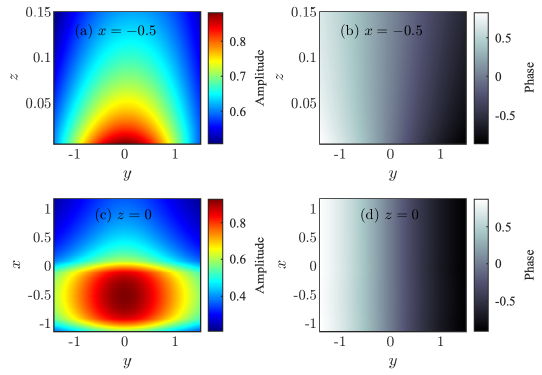}
    \caption{Scaling function $\Phi(x,y,z,0,0) = |\Phi|e^{i\psi_\Phi}$, describing the one-loop vertex correction in the quantum-critical regime (cf.\ \cref{eq:delta-g-scaling}). The functions $|\Phi|$ and $\psi_\Phi$ are plotted in $y,z$ plane for  $x=-0.5$ (a,b) and in (x,y) plane for $z=0$ (c,d).}
    \label{fig:vertex_qcp}
\end{figure}

In the quantum-critical regime the last two variables in \cref{eq:delta-g-scaling} can be set to zero.
The scaling function of the other three variables, $\Phi (x,y,z,0,0)$, is
\begin{multline}
    \Phi(x,y,z,0,0) = \frac{3^{3/4}}{4\pi}\int_{0}^{1}dr\int_{0}^{\infty}\frac{ds}{s^{3/2}+3^{3/4}\left|r+x\right|}\\
    \times \left[\frac{1}{(1-r)^{2/3}+r^{2/3}+i(y+z+2\sqrt{\left|z\right|s})} \right. \\
    \left. +\frac{1}{(1-r)^{2/3}+r^{2/3}+i(y+z-2\sqrt{\left|z\right|s})}\right].
\end{multline}
The scaling function $\Phi$ in \cref{eq:delta-g-scaling} is generally complex and can be parametrized as $\Phi = |\Phi|e^{i\psi_{\Phi}}$.
We plot
$|\Phi|$ and $\psi_{\Phi}$ in \cref{fig:vertex_qcp} for several values of parameters.
We see that in general $\Phi$ is of order one,
 but not particularly close to one, in distinction to the case $q=0$.
 In this situation, we expect that higher-order vertex corrections remain of order one, but do not change substantially
  the value of $\Delta g/g$ compared to one-loop result.

Finally, in the Fermi gas regime
$\omega^*_c < \omega_m < \omega_{\text{max}}$,
 we have for
 typical $\mathbf{q}$, and $\Omega_m \sim \omega_D^*$,
  $q_{\parallel}^{2}/(2m)/\tilde{\Sigma}(\omega_m) \ll 1$,~$(\omega^*_D)^2/(E_F {\tilde \Sigma}(\omega_m)) \ll 1$,~$v_{F}q_{\perp}/{\tilde \Sigma}(\omega_m) \sim 1$ and $\omega_m/{\tilde \Sigma} (\omega_m) \approx 1$.
  In this situation, we find $\Phi \sim \Sigma (\omega_m)/\omega_m \sim (\omega^*_c/\omega)^{1/3} \ll 1$, and hence
 $\Delta g (k, q)/g \ll 1$.

For qualitative analysis we do the same as for the electron-phonon case and estimate $\Delta g (k,q)/g$ by using
typical values of internal momenta and frequency. A simple experimentation shows that by order of magnitude
\begin{equation}
    \frac{\Delta g (k, q)}{g} \sim \frac{\Sigma (\omega_m)}{\tilde{\Sigma} (\omega_n) + v_F q^{\text{typ}}_{\perp} + \frac{(q^{\text{typ}}_{\parallel})^2}{2m}}
    \label{eq:aa12}
\end{equation}
In the Fermi liquid and Fermi gas regimes the vertex correction is small because either $(q^{\text{typ}}_{\parallel})^2/(2m) \gg \Sigma (\omega_m)$ or $\tilde{\Sigma} (\omega_n) \gg \Sigma (\omega_n)$.
In the quantum-critical, non-Fermi liquid regime all parameters in \cref{eq:aa12} are of the same order, and
$\Delta g (k, q)/g$ is generally of order one.

The quantitative measure of the strength of
the vertex correction is the magnitude of the two-loop contribution to the self-energy with the vertex correction included,
$\Sigma^{(2)}(\omega_m)$
compared to the one-loop $\Sigma^{(E)} (\omega_m)$.
We show the corresponding diagram in
\cref{fig:vertex_corr} (b). In the Fermi liquid and Fermi gas regimes, we find that
$\Sigma^{(2)}(\omega_m) \ll \Sigma^{(E)} (\omega_m)$,
consistent with the smallness of $\Delta g/g$.
In the quantum-critical regime, earlier order of magnitude studies~\cite{Altshuler1994,rech_2006,Khveshchenko_2006,Lee2009,metlitski2010quantum} have found that
$\Sigma^{(2)}(\omega_m)$
is of the same order as $\Sigma^{(E)}(\omega_m)$, unless one extends the theory to large $N$ or assumes that the prefactor for $q^2_{\parallel}$ in $\epsilon_{\mathbf{k}_F + \mathbf{q}}$ is much larger than $v_F/k_F$.

We confirmed this results, but went further and computed
$\Sigma^{(2)}(\omega_m)$ numerically
with the prefactor.
We
 found
\begin{equation}
    \Sigma^{(2)}(\omega_m)
    \simeq 0.038 \Sigma^{(E)}(\omega_m).
    \label{eq:sigma-2b}
\end{equation}
We see that while by order of magnitude
$\Sigma^{(2)}(\omega_m)$
is of order $\Sigma^{(E)}(\omega_m)$, it is far smaller numerically.
 It has been argued ~\cite{Lee2009,*Lee2018,metlitski2010quantum,Tobias2015,Eberlein_2016,pimenov_21} that
 self-energies  with  higher-loop order vertex corrections included  contain logarithmic singularities
We didn't compute these
 terms explicitly, but based on \cref{eq:sigma-2b} we expect them
 to contain small prefactors and
  remain small down to very small frequencies, before logarithmical singularities become relevant. We recall in this regard that the ground state near an Ising-nematic/Ising-ferromagnetic transition is a superconductor, hence in practice the behavior at the lowest frequencies is relevant only if for some reason superconductivity does not develop.

  For completeness, we also present results at weak coupling, when both $\lambda^*$ and $\lambda^*_E$ are small.
   There are two weak coupling regimes: $\lambda^* \ll (\lambda^*_E)^{1/2} \ll 1$ and $(\lambda^*_E)^{1/2} \ll \lambda^* \ll 1$. We focus on the second regime as it borders strong-coupling regime at $\lambda^* = O(1)$.
   In this  regime,   $\omega^*_c \ll  \omega^*$, i.e., there is no range of non-Fermi liquid behavior,
   although the self-energy still interpolates between $\lambda^* \omega_m$ at small frequencies and  $(\omega_m)^{2/3} (\omega_c^*)^{1/3}$ at higher frequencies (the relation between all characteristic frequencies is
   $\omega^*_c < \omega^* <\omega_D < \omega_{max}$).   A simple calculation shows that in this case
    vertex correction is small in $\lambda^*$ and typical $q^{\text{typ}}_{\perp}$ are parametrically larger than
    $q^{\text{typ}}_{\parallel}$ for all $\omega_m < \omega_{max}$ (see Fig. \ref{fig:pd_in}a for their values). Hence, Eliashberg theory
     is applicable and self-energy can be computed within perturbative one-loop approximation.

\begin{figure}
    \centering
    \includegraphics[width=\linewidth]{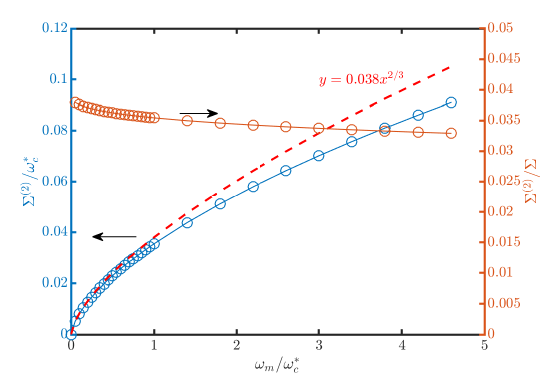}
    \caption{Two-loop fermionic self energy
        with vertex correction included, $\Sigma^{(2)} (\omega_m)$ (see \cref{fig:vertex_corr} (b)).
         Left axis: $\Sigma^{(2)} (\omega_m)$ in units of the characteristic frequency $\omega^*_c$ (the upper edge of non-Fermi liquid behavior).  Right axis: the ratio of $\Sigma^{(2)} (\omega_m)$ and the full one-loop Eliashberg self-energy $\Sigma (\omega_m) = \Sigma^{(E)} (\omega_m + \delta \Sigma (\omega_m)$ (see
          \cref{sec:subleading_in}).
        \label{fig:self_2lp_in}}
\end{figure}

\subsection{Summary of \cref{sec:qcm}}

Like for the electron-phonon system, there are three energy scales in the problem: the bosonic energy $\omega^*_D$, the coupling $g^*$ and the Fermi energy $E_F$.
This allows one to introduce two dimensionless ratios $\lambda^* = g^*/(4\pi \omega^*_D)$ and $\lambda^*_E = g^*/E_F$. The latter is a small parameter for the low-energy theory.
The strong coupling regime occurs at $\lambda^* >1$, $\lambda^*_E <1$.
In this regime, the system displays Fermi liquid behavior at $\omega < \omega^* \sim \omega^*_D \lambda^*_E/(\lambda^*)^2$, quantum-critical, non-Fermi liquid behavior with $\Sigma (\omega_m) \propto \omega_m^{2/3}$ at $\omega^* < \omega_m < \omega^*_c$, where
$\omega^*_c \sim g^* \lambda^*_E$,
and Fermi-gas behavior at larger frequencies.

The three key results for the system near an Ising-nematic/Ising-ferromagnetic  QCP are the following.

First, Eliashberg theory is rigorously justified at strong coupling $\lambda^* \gg 1$ in the Fermi liquid and Fermi gas regimes, but not in the quantum-critical regime.
In the latter, the leading vertex correction is $O(1)$ even when $\lambda^*_E$ is small. It is nevertheless small numerically,
  as evidenced by numerical smallness of the two-loop self-energy with vertex correction included.

Second, the small parameter
$\lambda_E^*$
simplifies the calculations within Eliashberg theory:
the self-energy $\Sigma^{(E)}(\mathbf{k}, \omega_m)$ is well approximated by the local $\Sigma^{(E)}(\omega_m)$, and the momentum integration in the computation of $\Sigma^{(E)}(\omega_m)$ can be factorized by invoking the Migdal ``fast electron/slow boson'' criterion.
This factorization is rigorously justified at all $\omega_m$ up to $\omega_{\text{max}} \sim \omega^*_c/(\lambda^*_E)^{3/2}
 \gg \omega^*_c$.
In this respect, $\lambda^*_E$ plays the same role as $\lambda_E$ for the electron-phonon case.

Third, and most important, the
 corrections to Eliashberg theory remain $O(1)$ (and numerically small) even at a QCP, where
$\omega^*_D =0$ and $\lambda^* = \infty$.
Furthermore, the ``fast electron/slow boson'' criterion also remains valid at a QCP as long as $\lambda^*_E$ is small.
 In other words, Eliashberg theory near a nematic QCP,
  while not rigorously justified,
 is quite accurate numerically and to the same numerical accuracy can be extended right to the QCP.\@

\begin{figure}
    \centering
    \includegraphics[width=\linewidth]{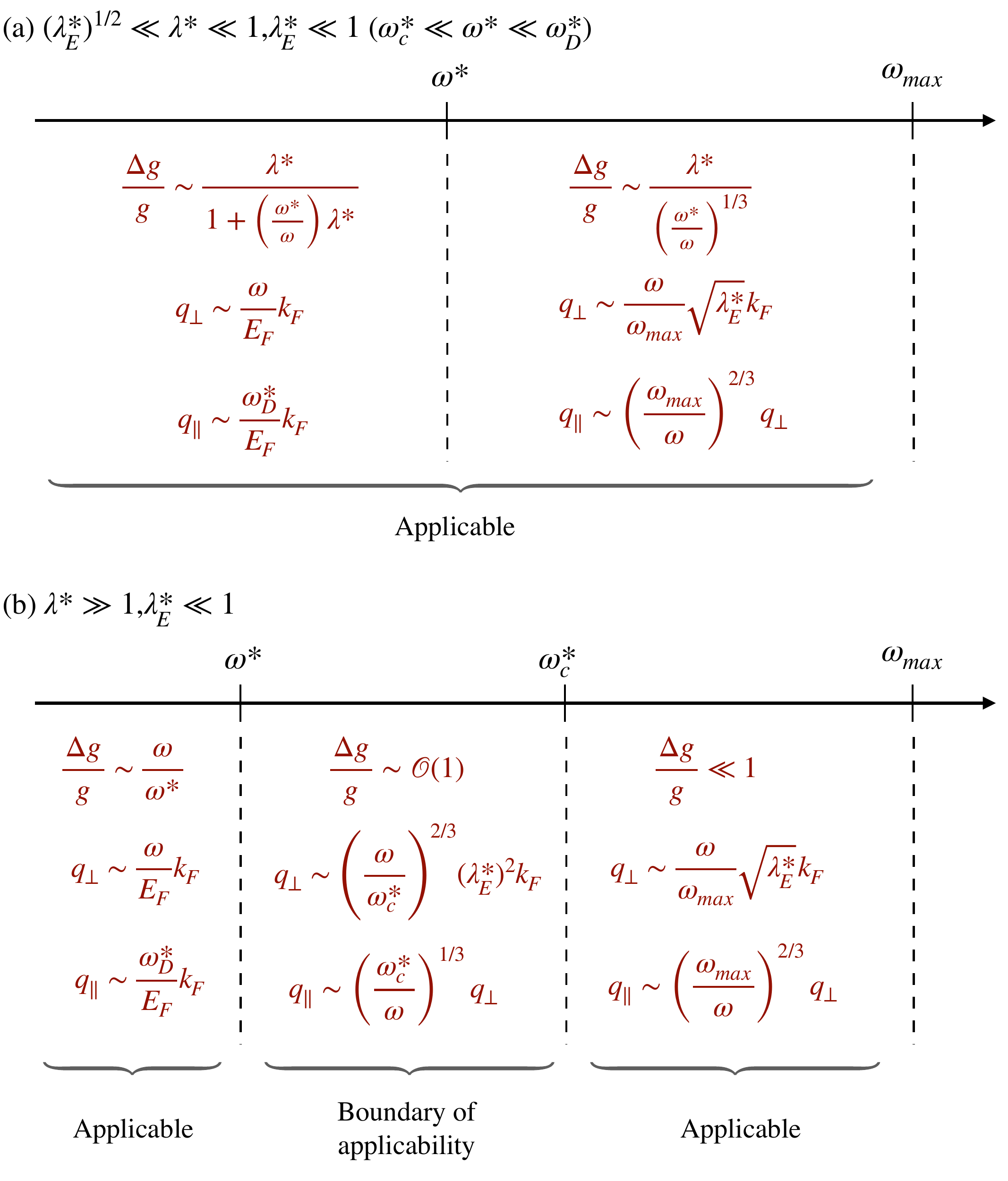}
    \caption{
    Illustration of the applicability of Eliashberg theory near Ising-nematic/Ising-ferromagnetic  critical point in a metal in (a) weak coupling regime at some distance from a critical point and (b) strong coupling regime near and at a critical point.
        \label{fig:pd_in}
    }
\end{figure}

We summarize the results for the Ising-nematic case in \cref{fig:pd_in}.

\section{Conclusions}
\label{sec:comp}

In this paper we compared the validity of Eliashberg theory for electrons interacting with an Einstein phonon and with soft nematic fluctuations near an Ising-nematic QCP and soft magnetic fluctuations near Ising-ferromagnetic QCP.\@
Eliashberg theory is the set of coupled one-loop self-consistent equations for the fermionic self-energy and polarization operator, with vertex corrections neglected.

For electron-phonon interaction, Eliashberg theory has been justified by the argument that an Einstein phonon is a slow mode compared to a fermion (the effective boson velocity is smaller than Fermi velocity). In the Ising-nematic/Ising-ferromagnetic  case this argument is not valid as soft fluctuations are collective modes of electrons and their velocity is of the same order as $v_F$.

We examined self-consistent Eliashberg theory and two-loop corrections to it for both cases in the strong coupling regime, where the system displays Fermi liquid behavior at the lowest energies, non-Fermi liquid behavior in a wide range of intermediate energies, and Fermi gas behavior at the largest energies.

For the electron-phonon case, Eliashberg theory is rigorously justified when the Eliashberg parameter $\lambda_E$ is small. Namely, the two-loop self-energy with vertex correction is small in $\lambda_E$.
Simultaneously, Eliashberg equations can be simplified using the same smallness of $\lambda_E$ to un-coupled perturbative one-loop equations, which can be easily solved.
In physical terms, the smallness of $\lambda_E$ follows directly from the condition that an Einstein phonon is a slow mode compared to electrons.

For the Ising-nematic/Ising-ferromagnetic case, soft bosons are collective modes of the electrons, and in the non-Fermi liquid quantum-critical regime there is no parametric smallness of the two-loop self-energy with vertex correction compared to the one-loop self-energy in the Eliashberg theory.
Nevertheless, we found that the two-loop self-energy is numerically much smaller than the one-loop one.
Additionally, the low-energy theory contains a small parameter $\lambda^*_E$, which plays the role of $\lambda_E$ in the sense that it again allows one to reduce the coupled self-consistent one-loop Eliashberg equations to un-coupled perturbative one-loop equations, which one can easily solve.
The implication of these results is that Eliashberg theory for the Ising-nematic/Ising-ferromagnetic case is on rather solid grounds.
 It is very likely that this
 holds also
  for other cases when fermions interact with their soft collective excitations in the charge or spin channel.

There is one aspect in which the Eliashberg description of fermions coupled to soft collective bosons works
even
better than for fermions interacting with an Einstein phonon.
Namely, for the collective boson case, the parameter $\lambda^*_E$ is independent of the distance to a QCP, and two-loop self-energy remains numerically smaller than the one-loop one even at a QCP.\@
As a result, Eliashberg theory can be extended right to the QCP.\@
For the electron-phonon case, the Eliashberg parameter $\lambda^*_E$ is inversely proportional to the dressed Debye frequency, and Eliashberg theory inevitably breaks down at some distance from the point where $\omega_D$ would vanish. Beyond this point, a completely new description in terms of polarons is needed~\cite{Alexandrov1994,Millis_1996,x_representation,sous2023bipolaronic,PhysRevX.13.011010}.

\begin{acknowledgments}
We thank Ar.~Abanov, E.~Berg, H.~Goldman, R.~Fernandes, M. Foster, I.~Esterlis, P.A.~Lee, D.~Maslov, A.~Millis, C.~Murthy, S.~Kivelson, A.~Klein, P. Nosov, N.~Prokofiev, S. Raghu, B.~Svistunov, S.~Sachdev, J.~Schmalian, Y. Wang  and Y-M. Wu for helpful discussions.
The work of A.Ch.\ was supported by U.S.~Department of Energy, Office of Science, Basic Energy Sciences, under Award No.~DE-SC0014402.
Part of the work was done at the Kavli Institute for Theoretical Physics (KITP) in Santa Barbara, CA\@.
KITP is supported by the National Science Foundation under Grants No.~NSF PHY-1748958 and PHY-2309135.
\end{acknowledgments}

\bibliography{ref.bib}

\end{document}